\documentclass[twocolumn,showpacs,preprintnumbers,nofootinbib,prd,
superscriptaddress,10pt]{revtex4-1}
\usepackage{lmodern}
\usepackage{amsmath,amssymb}
\usepackage[normalem]{ulem}
\usepackage{textcomp}
\usepackage{hyperref}
\usepackage{bm}
\usepackage{graphicx}
\usepackage{psfrag}
\usepackage[usenames,dvipsnames]{xcolor}
\usepackage[utf8]{inputenc}
\usepackage{lipsum}
\usepackage{aas_macros}
\usepackage{booktabs}
\usepackage{setspace}
\usepackage{array}
\usepackage{multirow}
\usepackage{orcidlink}

\hypersetup{
    bookmarks=true,         
    unicode=false,          
    pdftoolbar=true,        
    pdfmenubar=true,        
    pdffitwindow=false,     
    pdfstartview={FitH},    
	pdftitle={Large mass ratio IMR eccentric model},
    pdfauthor={Pratul Manna et al.},
    pdfsubject={Gravitational Waves},   
    pdfkeywords={gravitational waves, eccentric compact binaries, general 
		relativity},
    pdfnewwindow=true,      
    colorlinks=true,       
    linkcolor=red,          
    citecolor=Blue,        
    filecolor=magenta,      
    urlcolor=Blue,           
    linktocpage=true
}

\allowdisplaybreaks[4]

\DeclareFontFamily{OT1}{pzc}{}
\DeclareFontShape{OT1}{pzc}{m}{it}{<-> s * [1.10] pzcmi7t}{}
\DeclareMathAlphabet{\mathpzc}{OT1}{pzc}{m}{it}

\newcommand{\hlm}{\mathpzc{h}_{\ell m}}

\graphicspath{{./figures/}}

\newcommand{\cmmnt}[1]{}

\newcolumntype{C}[1]{>{\centering\arraybackslash}p{#1}}

\begin{document}

\title{Large mass-ratio ($3< q \leq 10$) inspiral-merger-ringdown model for eccentric binary black holes}
\date{\today}

\author{Pratul Manna \orcidlink{0009-0006-4918-2943}}
\email{mpratul@astrouw.edu.pl}
\affiliation{Astronomical Observatory, University of Warsaw, Al. Ujazdowskie 4, 00478 Warszawa, Poland}
\affiliation{Department of Physics, Indian Institute of Technology Madras, Chennai 600036, India}
\author{Tamal RoyChowdhury \orcidlink{0000-0003-4242-0532}}
\email{tamal@uwm.edu}
\affiliation{University of Wisconsin Milwaukee, Milwaukee, WI 53201, USA}
\affiliation{Department of Physics, Indian Institute of Technology Madras, Chennai 600036, India}
\author{Chandra Kant Mishra \orcidlink{0000-0002-8115-8728}}
\email{ckm@physics.iitm.ac.in}
\affiliation{Department of Physics, Indian Institute of Technology Madras, Chennai 600036, India}
\affiliation{Centre for Strings, Gravitation and Cosmology, Department of Physics, Indian Institute of Technology Madras, Chennai 600036, India}
\begin{abstract}
     We present \texttt{EccHybIMR22}, a fully analytical \textbf{Ecc}entric-\textbf{Hyb}rid-\textbf{I}nspiral-\textbf{M}erger-\textbf{R}ingdown model for the quadrupole ($\ell$=$2$, $|m|$=$2$ or simply the \textbf{22}) mode. The model is obtained by smoothly attaching an eccentric-inspiral to a quasi-circular-merger-ringdown prescription and is trained on eccentric numerical relativity (NR) simulations with mass ratios $q\leq 10$. In particular, we focus on obtaining a model that is sensitive to large mass ratios ($q>3$). Leveraging the recent expansion of SXS catalog of NR simulations, we construct a set of 91 hybrids (obtained by combining waveforms from post-Newtonian (PN) theory with NR simulations) with a target low total mass limit ($M_{\rm low}$) of $20M_{\odot}$ at an averaged 22 mode frequency of 20Hz, inaccessible with existing NR simulations alone. Employing a little less than a third of these hybrids as targets, we construct \texttt{EccHybIMR22} and recover 65\% of the hybrids from the remaining set with mismatches smaller than $\sim3\%$ for a total mass of $40M_\odot$ (chosen arbitrarily) and orbital eccentricity $e_{\rm 20Hz}\lesssim0.33$ ($e_{\rm 20Hz}\lesssim 0.12$) for mass ratios $q\leq3$ ($q>3$). Furthermore, when tested  against 175 SXS simulations, the model is able to recover nearly $80\%$ of the simulations with similar mismatches for systems with $0\lesssim e^{\rm NR}_{\rm 20Hz}\lesssim 0.3$. 
\end{abstract}

\maketitle
\section{Introduction}
\label{sec:intro}

\noindent 
Orbital eccentricity offers unique insights into the astrophysical processes and environments governing the formation and evolution of compact binary coalescences~\cite{Mapelli:2021taw}. Over a decade since the first detection of gravitational waves (GWs)~\cite{Abbott:2016blz}, the LIGO-Virgo-KAGRA (LVK) collaboration has reported $390$ significant events observed until the completion of the second part of the fourth observing run (O4b)~\cite{LIGOScientific:2026wfs, LIGOScientific:2026sit, LIGOScientific:2026ifv, LIGOScientific:2020ibl, LIGOScientific:2021usb, LIGOScientific:2021djp, LIGOScientific:2025slb} as part of the Gravitational Wave Transient Catalogs (GWTC) and yet none of the events presented unambiguous signatures of the presence of orbital eccentricity. 
And thus, even though these observations have revolutionised our understanding in the fields of astrophysics, cosmology and fundamental physics, the formation channels of these sources still remain uncertain~\cite{Fumagalli:2024gko, Antonini:2024het, Gupte:2026whi}.\\

Binaries formed in isolated environments efficiently circularize by dissipating eccentricity via radiation reaction forces~\cite{Peters:1964zz}. However, in dynamical environments like globular clusters~\cite{Samsing:2017xmd,Samsing:2013kua}, field triples~\cite{Kozai:1962zz,Lidov62} or active galactic nuclei~\cite{Tagawa:2020jnc,Grishin:2025ofa}, approximately $5-10\%$ of the total binary population can retain detectable eccentricity \cite{Zevin:2018kzq, Rodriguez:2017pec, Ramos-Buades:2020eju, Fumagalli:2024gko, Tagawa:2020jnc,LIGOScientific:2019dag, LIGOScientific:2023lpe,Singh:2025ojp, Rozner:2026jtj, Grishin:2025ofa,Dorozsmai:2025jlu,Chattopadhyay:2023pil, Pompili:2026yxq} while entering the sensitivity bands of ground based detectors such as LIGO \cite{Capote:2024rmo} and Virgo \cite{Acernese:2015gua}. Quasi-circular (QC) templates are not efficient in detecting eccentric signals in template-based searches~\cite{Favata:2021vhw,DePorzio:2024cet,Lenon:2021zac,Brown:2009ng,Huerta:2013qb,Phukon:2024amh, Phukon:2025cky}. Furthermore, neglecting eccentricity can lead to biased parameter inferences (in particular, of chirp mass) \cite{Favata:2021vhw, Abbott:2016wiq, Divyajyoti:2023rht, Das:2024zib,Huez:2025npe, RoyChowdhury:2026xgb} motivating accurate modelling of eccentricity related effects. Furthermore, third-generation (3G) detectors like Einstein Telescope \cite{Punturo:2010zz, Hild:2010id} and Cosmic explorer \cite{McClelland:T1500290-v3, Dwyer:2014fpa, Evans:2016mbw} are more likely to detect eccentricity due to their sensitivity to low frequencies and thus detection and parameter estimation analyses will inevitably require better handling of eccentricity at the level of waveforms in the 3G era. Moreover, this requirement grows as we move from ground-based to proposed/planned space-based experiments such as DECIGO~\cite{Kawamura:2020pcg} and Laser Interferometer Space Antenna (LISA)~\cite{LISA:2024hlh} with target sensitivities in the deci-Hz (dHz) and mili-Hz (mHz) band, respectively.\\

Even though LVK collaboration did not report eccentricity signatures in the analyzed events \cite{LIGOScientific:2019dag, LIGOScientific:2023lpe, LIGOScientific:2026wfs}, there have been studies that highlight possible candidates among the observed population. For instance, Refs.~\cite{Romero-Shaw:2020aaj,Romero-Shaw:2019itr,Romero-Shaw:2021ual,Romero-Shaw:2022xko,Gupte:2024jfe,Planas:2025jny} show evidence of eccentricity in following binary black hole (BBH) events -- GW$190620$, GW$191109$,
GW$200208\_22$, GW$190701$, GW$200129$, GW$200208\_22$, GW$200129$, GW$200208\_22$, GW$190701$ and GW$190929$. Additionally, the BNS event GW$200105$ was also found to be eccentric as Refs.~\cite{Planas:2025plq, Morras:2025xfu} report. Furthermore, Refs.~\cite{Gamba:2021gap,Gayathri:2020coq,Romero-Shaw:2020thy,Jan:2026mnp} find eccentricity in events consisting of intermediate mass black holes, such as GW$190521$~\cite{Abbott:2020tfl}, although, there are also studies that refute similar claims~\cite{Gupte:2024jfe, Iglesias:2022xfc, Ramos-Buades:2023yhy, Gamboa:2024hli}. Clearly, even if there is, the evidence for eccentricity is not strong enough. This (at least) in part can be attributed to absence of accurate eccentric waveform models as the first generation employable models only recently became available~\cite{Ramos-Buades:2026kbq, Paul:2024ujx, Gamboa:2026jht, Gamboa:2026ybr}.\\ 

Several efforts aimed at improving eccentric waveform models (including spin-effects for some cases) based on Post-Newtonian (PN) methods \cite{Islam:2025oiv, Gopakumar:1997bs, Gopakumar:2001dy, Laxman:2026xts, Paul:2022xfy, Mishra:2015bqa,Boetzel:2019nfw,Ebersold:2019kdc,Sridhar:2024zms,Henry:2023tka,Arun:2007rg,Arun:2007sg,Tanay:2016zog, Trestini:2025yyc}, Effective-One-Body (EOB) framework \cite{Gamboa:2024hli, Gamboa:2026jht, Gamboa:2026ybr, Ramos-Buades:2021adz, Gamba:2024cvy, Nagar:2024oyk, Chiaramello:2025bhi,Placidi:2021rkh,Placidi:2023ofj,Albanesi:2022xge,Albanesi:2025txj,Nagar:2024dzj}, phenomenological techniques \cite{Planas:2025feq,Ramos-Buades:2026kbq,Islam:2024bza,Manna:2024ycx,Chattaraj:2022tay,Paul:2024ujx,Huerta:2017kez,Hinder:2017sxy,Islam:2024tcs} and surrogate approaches \cite{Islam:2024zqo, Maurya:2025shc, Nee:2025nmh,Varma:2019csw} are underway.\\

Construction of inspiral-merger-ringdown (IMR) models mostly relies on training the model using numerical relativity (NR) simulations. In the light of recent update to the SXS catalog, this paper focuses on enhancing the sensitivity of the waveform model presented in Refs.~\cite{Chattaraj:2022tay, Manna:2024ycx} to large mass-ratios ($3<q\leq 10$), as opposed to the range $1\leq q\leq3$ explored earlier. This is particularly important since binaries with high mass ratios can retain detectable eccentricity owing to their slow inspiral rate and also bridge the gap towards intermediate and extreme mass ratio inspirals (IMRIs and EMRIs) which are key targets for space-based detectors in deci-Hz~\cite{Kawamura:2020pcg, Sharma:2026tff} and mili-Hz band~\cite{LISA:2024hlh, Babak:2017tow}, respectively. For constructing such a model we broadly follow the scheme presented in \cite{Chattaraj:2022tay, Manna:2024ycx}. First, PN waveforms are compared against an expanded set of 149 NR simulations (with reference eccentricity $e_{\rm ref}^{\rm NR}\in [0.01,0.8]$) from the updated SXS catalog~\cite{Scheel:2025jct}.\footnote{The reference eccentricity values for each simulation is obtained from the simulation metadata.} Then, following the method of  Ref.~\cite{Manna:2024ycx} for hybridization, the PN and NR waveforms are matched in an appropriate time-window to construct 91 hybrids; 24 of which are then used to construct \texttt{EccHybIMR22}. The model (\texttt{EccHybIMR22}) is subsequently validated against another set of 64 hybrids and 175 NR simulations.\footnote{Note that, the remaining three hybrids are excluded from \textit{testing set} as they have reference eccentricities larger than the ones explored in this work.}\\

The rest of the article is structured as follows: Section~\ref{sec:summary} provides an executive summary of the current work. Section~\ref{sec:methods} discusses the criteria and construction of hybrids. Our analytical model is presented in Section~\ref{sec:22-mode-model} and is validated against a set of hybrids (not employed in training the model) as well as against NR simulations in Sec.~\ref{sec:validation}. Finally, in Sec.~\ref{sec:disc}, we summarise our findings and conclude the paper. Technical details of hybrids as well as of the model together with the waveform visualisations are pushed to various appendices to help legibility.

\section{Executive Summary}
\label{sec:summary}

The development of complete inspiral-merger-ringdown family of waveforms relies heavily on PN methods for the early inspiral phase of the signal and on NR simulations for the late inspiral, merger, ringdown phase. For instance, phenomenological models (including the kinds developed in our earlier works~\cite{Chattaraj:2022tay,Manna:2024ycx}) construct a set of PN-NR hybrids by matching the waveforms from the two approaches in a time-window where both are expected to agree with each other. Subsequently, a fully analytical model, trained on these hybrids is constructed by finding appropriate fits. Waveforms constructed here broadly follow the above scheme. Our inspiral model combines the 3PN accurate dominant mode amplitude model based on Refs.~\cite{Mishra:2015bqa,Boetzel:2019nfw,Ebersold:2019kdc} with the 2PN accurate phasing model presented in Ref.~\cite{Tanay:2016zog}.\footnote{The time-domain version of the model of Ref.~\cite{Tanay:2016zog} is named, \texttt{EccentricTD}.} Both the amplitude and the phase account for corrections in eccentricity to the $6^{\rm th}$ power. This inspiral prescription is compared with a set of nonspinning eccentric BBH simulations produced using the Spectral Einstein Code (SpEC)~\cite{Scheel:2025jct} in a specific time-window. The similarity between PN and NR wavefroms within a specified time window is quantified using the match ($\mathcal{M}$) which is defined as~\cite{Purrer:2014fza},
\begin{subequations}
\begin{equation}
    \mathcal{M}(h_1,h_2)
    =
    \max_{\phi_c,t_c}
    \frac{
    \left\langle e^{i(\phi_c+2\pi f t_c)}h_1 \mid h_2 \right\rangle
    }{
    \sqrt{\langle h_1|h_1\rangle \langle h_2|h_2\rangle}
    },
\end{equation}
with,
\begin{equation}
    (h_1|h_2) \equiv 4\ {\rm Re}\left[
    \int_{0}^{\infty}
    \frac{\tilde{h}_1(f)\tilde{h}_2^{*}(f)}
    {S_n(f)}
    \,\mathrm{d}f
    \right].
\end{equation}
\label{eq:match}
\end{subequations}

\noindent where ($\tilde{h}_1,\tilde{h}_2$) denote the Fourier transforms of ($h_1,h_2$), $S_n(f)$ represents the one-sided power spectral density of the noise of the detector.\\

Figure~\ref{fig:PN-NR_scatter} shows the mismatch ($1-\mathcal{M}$) between the PN and NR waveforms, evaluated over the optimal time-window (providing the maximum $\mathcal{M}$) as a function of the reference NR eccentricity ($e_{\rm ref}^{\rm NR})$. 
NR simulations with  $e_{\rm ref}^{\rm NR}\lesssim0.45$ provide $\mathcal{M}\gtrsim96\%$ which is used as the criterion to construct the hybrids. 
These hybrids are subsequently used to construct a numerical model by attaching a PN model  
for the inspiral part and \texttt{SEOBNRv5} \cite{Pompili:2023tna} for the QC merger-ringdown.\\ 
 
The numerical fits are then calibrated against a particular set of 24 hybrids labelled as \textit{training set} to build the analytical model we refer to as \texttt{EccHybIMR22}.
The model is then validated against the remaining 64 hybrids (\textit{testing set}). Details of all hybrids are summarised in Table~\ref{tab:nrsims}. Figure~\ref{scatter_comp_training_testing} displays the mismatches of the model against the target hybrids for a total mass of $40M_\odot$; see also the strain comparison plots of Fig.~\ref{strain_comp_training} and Fig.~\ref{strain_comp_testing}. The model recovers 44 of the 64 hybrids from the \textit{testing} set with matches $\gtrsim96.5\%$ for eccentricities $e_0\lesssim0.33$ ($e_0\lesssim0.12$), and mass ratios $q\leq3$ ($q>3$).\\  

Another check of the performance of our model comes from its comparison against 175 NR simulations (including circular simulations) with $e_{\rm ref}^{\rm NR}\in[0,0.45]$ from the SXS catalog. Figure~\ref{fig:mismatch_NR} shows the mismatches (as functions of $q$ and $e_{\rm ref}^{\rm NR}$) for the lowest total mass that can be probed with each NR simulation. We recover 140 of the 175 simulations with matches $\gtrsim96.5\%$ for systems in the eccentricity range $0\lesssim e_{\rm ref}^{\rm NR}\lesssim0.3$ (see middle panel of Fig.~\ref{fig:mismatch_NR}). Compared to the mismatches obtained using the model presented in Ref.~\cite{Manna:2024ycx} (referred to as `Old Model' throughout the paper) we find that, \texttt{EccHybIMR22} performs better in high $-\ q$ and high $-\ e_{\rm ref}^{\rm NR}$ regimes; see left panel of Fig.~\ref{fig:mismatch_NR}. For $e_{\rm ref}^{\rm NR}\gtrsim0.1$ and $q>3$, \texttt{EccHybIMR22} provides higher (averaged) matches (up to $\sim43\%$) compared to the old model. We also compare an alternate model based on artificial neural network (ANN)~\cite{paszke2019pytorch} which recovers a comparable fraction of NR simulations; see the right panel of Fig.~\ref{fig:mismatch_NR}. Finally, Fig.~\ref{ecc_hist} explores the adequacy of the choice to attach a QC merger-ringdown prescription in our model. As can be seen there, the model produces sub-threshold match ($\mathcal{M}\lesssim96.5\%$) for $e_{\rm ref}^{\rm NR}\gtrsim0.15$ hinting at the possible inappropriateness of the approximation for larger eccentricities.\\

\section{The Hybrid Model}
\label{sec:methods}

\begin{figure*}[htbp!]
    \centering
    \includegraphics[width=
    \textwidth]{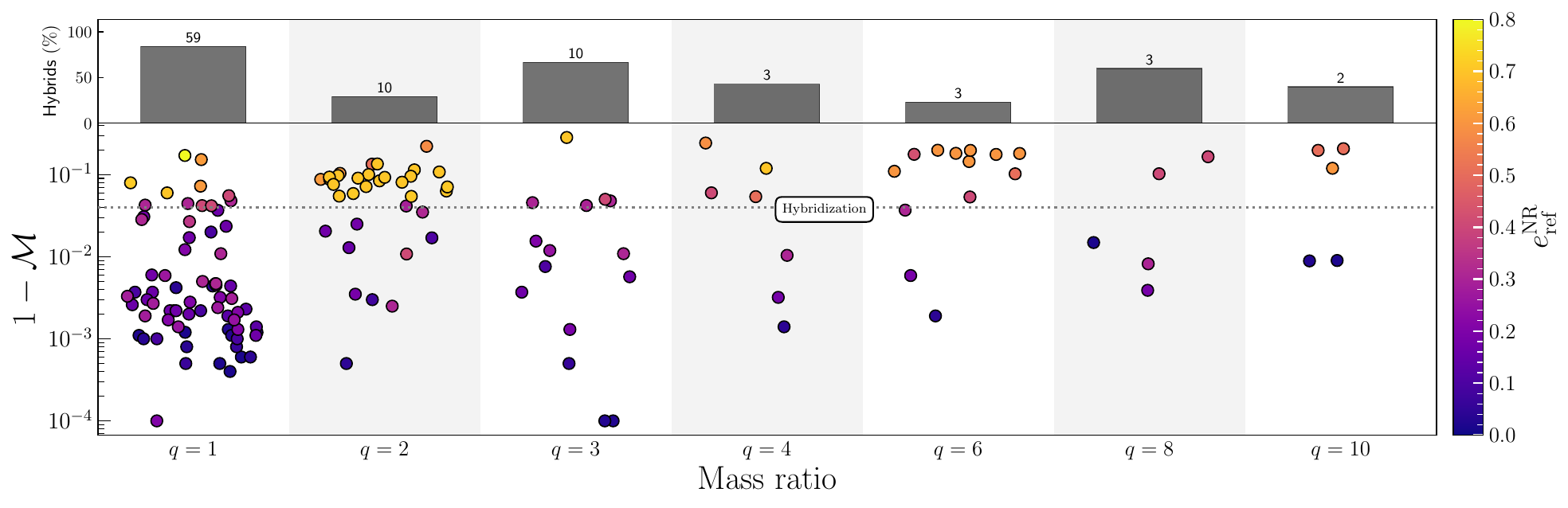}
    \caption{\textit{Bottom panel:} Displays mismatches ($1-\mathcal{M}$; see Eq.~\eqref{eq:match}) between PN and NR waveforms in a time-window where the two agree the most. The colorbar reports the reference value of eccentricity ($e^{\rm NR}_{\rm ref}$) obtained from the metadata of the corresponding NR simulation. The dotted horizontal line shows $96\%$ match (or equivalently 4\% mismatch) used as the cut-off value for hybridization. \textit{Top panel:} Shows the percentage (and the number) of hybridizations within each mass ratio; a total of 91 hybrids are constructed. The simulations are arranged based on ascending mass-ratio from left to right. {Note that, the only $q=5$ hybrid is not shown here but is listed in Table~\ref{tab:nrsims}.}}
    \label{fig:PN-NR_scatter}
\end{figure*}

\begin{figure}[htbp!]
    \includegraphics[width=0.49
    \textwidth]{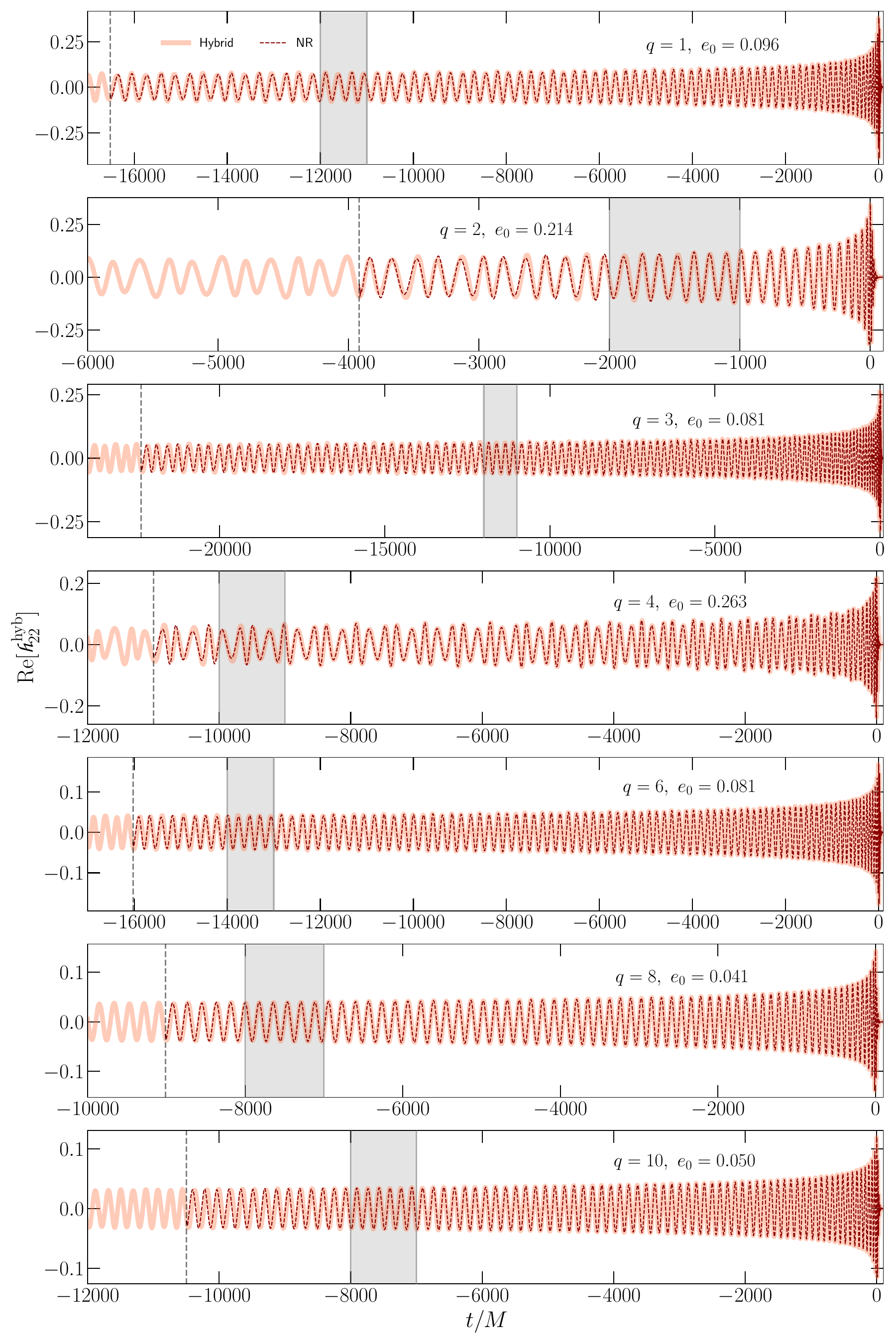}
    \caption{Hybrid waveforms for seven distinct mass ratios constructed by matching a NR simulation with a PN inspiral waveform within the optimal time window $t \in (t_{\rm w}, t_{\rm w}+1000M)$ denoted by shaded grey region.  
    The dashed vertical line denotes $t_0^{\rm NR}$ (start time of the NR simulation). The mass ratio ($q$) and initial eccentricity ($e_0$) for each hybrid are also reported. All the 90 hybrids are listed in Table~\ref{tab:nrsims}.
    } 
    
    \label{fig:hybridplot}
\end{figure}

Construction of the hybrids start with comparing PN and NR prescriptions. For each NR simulation, we iterate over several time-intervals of duration $1000M$ between the start time of the simulation ($t_0^{\rm NR}$) and a time $-2000M$ prior to the merger (set to $t=0$). For each such interval, we calculate the match ($\mathcal{M}$) between the PN and NR waveforms and the time-window providing the highest match is chosen for hybridization provided the match is better than $96\%$. We find that simulations with $e^{\rm NR}_{\rm ref}\lesssim 0.45$ are suitable for hybridization based on the above criterion. Figure~\ref{fig:PN-NR_scatter} displays the mismatch ($1-\mathcal{M}$) between the PN and NR waveforms in the hybridization window for all 149 simulations. 91 of these simulations have matches better than 96\% in the hybridization window; these are the simulations below the dotted line labelled as ``Hybridization" in Fig.~\ref{fig:PN-NR_scatter}. We also note that nearly all simulations above the dotted line have large eccentricities and sub-threshold matches ($<96\%$). This is likely due to inadequacy of PN models to probe such eccentricities. See for instance \cite{Ramos-Buades:2026kbq} which makes an effort to push the validity of inspiral model by accounting for higher order eccentricity corrections. Further, we also extract using the {\fontfamily{qcr}\selectfont{gw\_eccentricity}} package~\cite{Shaikh:2023ypz, Shaikh:2025tae} reference values of orbital eccentricity and mean anomaly at a given averaged 22 mode frequency identified as the trio --- $(e^{\rm gw}_{\rm ref},l^{\rm gw}_{\rm ref},x^{\rm gw}_{\rm ref})$. The PN frequency parameter $x$ is defined as $x_{\rm gw}=(\pi Mf_{\rm gw})^{2/3}$ where $M$ and $f_{\rm gw}$ represent the total mass and GW frequency respectively. These together with the start time of the hybridization window ($t_{\rm w}$) and other relevant parameters for each simulation are listed in Table~\ref{tab:nrsims}. (SXS simulation IDs have been retained to identify the hybrids with the corresponding NR simulation.)\\

Once the hybridization window is identified, the hybrids are constructed following the procedure described in Refs.~\cite{Chattaraj:2022tay, Manna:2024ycx, Varma:2016dnf, Varma:2014jxa} for hybridization. The first steps involve least square minimization of the integrated difference between the PN and NR waveforms over the desired time-interval. Formally, we construct the following

\begin{equation}
\begin{aligned}
\delta = \min_{t_0,\varphi_0, \psi} \!\!\! \int\limits_{t_{\rm w}}^{t_{\rm w} +1000M} \!\!\! dt \sum_{\ell,m} \left|\hlm^\mathrm{NR}(t-t_0) \, e^{i(m\varphi_0+\psi)} -\hlm^\mathrm{PN}(t)\right|,
\label{eq:delta}
\end{aligned}
\end{equation}

\noindent
where the minimization is performed over a time shift ($t_0$) and the two angles ($\varphi_0, \psi$).\\ 

The hybrid waveforms are subsequently constructed by combining the NR data with the ``best matched'' PN waveform in the following way:

\begin{equation}
\hlm^\mathrm{hyb}(t) \equiv \, \tau_1(t) \, \hlm^\mathrm{NR}(t-t_0') \ e^{i(m\varphi_0'+\psi')} + (1-\tau_1(t)) \, \hlm^\mathrm{PN}(t) ,
\label{eq:hyb}
\end{equation}

\noindent
where ($t_0'$, $\varphi_0'$, $\psi'$) are the values of ($t_0$, $\varphi_0$, $\psi$) that minimize the integral of Eq.~\eqref{eq:delta}. In the above equation, $\tau_1(t)$ is a weighting function defined by

\begin{eqnarray}
\tau_1(t) \equiv \left\{ \begin{array}{ll}
0 & \textrm{if $t < t_{\rm w} $}\\ \\ 
\frac{t-t_{\rm {w}}}{1000M}  & \textrm{if $t_{\rm{w}} \leq t < t_{\rm{w}}+1000M $}\\ \\
1 & \textrm{if $t_{\rm{w}}+1000M \leq t$}
\end{array} \right.
\label{eq:tau}
\end{eqnarray}

The hybrids corresponding to a set of representative NR simulations with seven distinct mass-ratios for the dominant mode are shown in Fig.~\ref{fig:hybridplot}. For comparison we also plot corresponding NR simulations.  
The two are aligned at merger and the shaded grey region $t \in (t_{\rm w}, t_{\rm w}+1000M)$ highlights the matching window where hybridization is performed.\footnote{Note that both Ref.~\cite{Chattaraj:2022tay, Manna:2024ycx} perform hybridization in a fixed time-window of $t\in [-1000M, -2000M]$ for all simulations considered there.} Overlapping hybrid and NR waveforms outside on the left of the matching window hint at the quality of hybridization performed here. 
Each hybrid starts at a fixed initial frequency $x_0=0.0337$ with an initial eccentricity $(e_0)$ and a mean anomaly ($l_0$) such that it attains the parameter values $(e^{\rm gw}_{\rm ref},l^{\rm gw}_{\rm ref})$ at the reference frequency $(x^{\rm gw}_{\rm ref})$.
Assuming advanced LIGO sensitivity~\cite{TheLIGOScientific:2014jea}, the choice of $x_0=0.0337$ corresponds to a total mass of $20M_\odot$ at the averaged 22 mode frequency of $20{\rm Hz}$, and is the lowest total mass ($M_{\rm low}$) we can explore using our hybrids. While hybrids can be generated for arbitrary $x_0$ since the PN model is analytic, soon it may become inadequate due to its poor sensitivity to large initial eccentricities, making such hybrids less reliable at low frequencies as also remarked at the beginning of this section.\\

\section{The Analytic Model}
\label{sec:22-mode-model}

In this section we present our analytical model which is obtained by attaching an eccentric PN inspiral to a QC merger-ringdown model. For the amplitude part of the inspiral model, unlike Ref.~\cite{Chattaraj:2022tay, Manna:2024ycx} which utilizes \texttt{EccentricTD}~\cite{Tanay:2016zog} that is only Newtonian accurate in amplitude, we employ 3PN accurate amplitude expressions from Refs.~\cite{Mishra:2015bqa,Boetzel:2019nfw,Ebersold:2019kdc}. However, we choose to continue with \texttt{EccentricTD} for the phasing which is 2PN accurate. Keeping the amplitude and phase same as the hybrids reduces any unknown systematics between the inspiral parts of target and the model. For the merger-ringdown portion we use the EOB based QC model \texttt{SEOBNRv5}~\cite{Pompili:2023tna}. The model is obtained in two steps. First, a numerical version of the model is obtained by iterating simultaneously over a time shift ($t_{\rm shift}$) and attachment times for amplitude and frequency ($t_{\rm match}^{\mathcal{A}},t_{\rm match}^{\omega}$).\footnote{Note that, the optimization is performed over the frequency of the numerical model against the hybrid. Thereafter, the phase is obtained by integrating the frequency vector.} 
The second step involves obtaining analytical fits for these and are eventually used to generate the model.\\  

The procedure for constructing the numerical model mostly follows Refs.~\cite{Manna:2024ycx,Chattaraj:2022tay}. Since the merger time is unknown, we choose it to be the time of peak amplitude for the circular IMR template and set it to zero. Next, a time shift ($t_{\rm shift}$) is applied on the PN inspiral relative to the circular waveform. The amplitude and frequency model is obtained by joining the eccentric inspiral with the circular template over a certain time window. The general structure of the amplitude ($\mathcal{A}_{22}$) and frequency ($\omega_{22}$) models are shown below,
{
\begin{equation}
\begin{aligned}
(\mathcal{A},\omega)_{22}^\mathrm{model}(t)
\equiv {} & \tau_2(t)\,(\mathcal{A},\omega)_{22}^\mathrm{IMR}(t) \\
& + \bigl(1-\tau_2(t)\bigr)\,(\mathcal{A},\omega)_{22}^\mathrm{inspiral}(t)
\end{aligned}
\label{eq:amp_model}
\end{equation}

where $\tau_2(t)$ is defined as 

\begin{eqnarray}
\tau_2(t) \equiv \left\{ \begin{array}{ll}
0 & \textrm{if $t < t_{\rm{i}} $}\\
\frac{t-t_{\rm{i}}}{t_{\rm{f}}-t_{\rm{i}}}  & \textrm{if $t_{\rm{i}} \leq t < t_{\rm{f}} $}\\
1 & \textrm{if $t_{\rm{f}} \leq t$.}
\end{array} \right.
\label{eq:tau_amp}
\end{eqnarray}
For amplitude (frequency) model, $t_{\rm i}=t_{\rm{match}}^\mathcal{A}$ ($t_{\rm{match}}^\omega$) and $t_{\rm f} \approx-30M$, where $t_{\rm match}^{\mathcal{A}}$ and $t_{\rm match}^{\omega}$ denote the attachment times of amplitude and frequency respectively. 
Unlike Refs.~\cite{Manna:2024ycx,Chattaraj:2022tay}, we simultaneously vary $(t_{\rm shift},t_{\rm match}^{\mathcal{A}},t_{\rm match}^{\omega})$ trio and optimize the numerical model against the hybrids 
using the Nelder-Mead downhill simplex minimization algorithm in \textsc{Scipy} \cite{scipy2020}. We obtain numerical fits for all target hybrids using the same process (See also Appendix~\ref{sec:appn_fits} for details).\\

Nearly $24$ of the hybrids (with matches $\gtrsim96\%$ with the numerical model) representing seven distinct mass ratios and including at least 3 simulations for each $q$ value are chosen for training the model and referred to as `\textit{training set}'.\footnote{For mass ratios with fewer simulations (particularly for large mass ratios, likely due to higher computational cost)  we try dividing number of simulations between the two sets as evenly as possible to represent different mass ratios.} The remaining ones (referred to as `\textit{testing set}') are used to validate \texttt{EccHybIMR22}. Note that, both sets of hybrids broadly represent the entire mass-ratio space ($q\in [1,10]$) we wish to explore in this work. Note also that, the model is trained on hybrids with a fixed total mass of 30$M_{\odot}$. While this could be as low as 20$M_{\odot}$, we make a conservative choice of 30$M_{\odot}$ when constructing fits to the training set hybrids. This choice does not have any impact on the models that are trained on these hybrids, instead, it guarantees better accuracy (lower $e_{\rm 20Hz}$) for the inspiral portion of the training set hybrids.\\

Next, in order to generate the model for arbitrary parameter configurations within the parameter space of calibration, we develop analytical fits for the $(t_{\rm shift},t_{\rm match}^{\mathcal{A}},t_{\rm match}^{\omega})$ trio as functions of mass-ratio, eccentricity and mean-anomaly of the binary system. 
The analytical expressions and the fitted coefficients (see Tables~\ref{tab:tshift-table} - \ref{tab:tfreq-table}) are provided in Appendix~\ref{sec:appn_fits}.

\begin{figure*}[htbp!]
\centering
    \includegraphics[width=\textwidth]{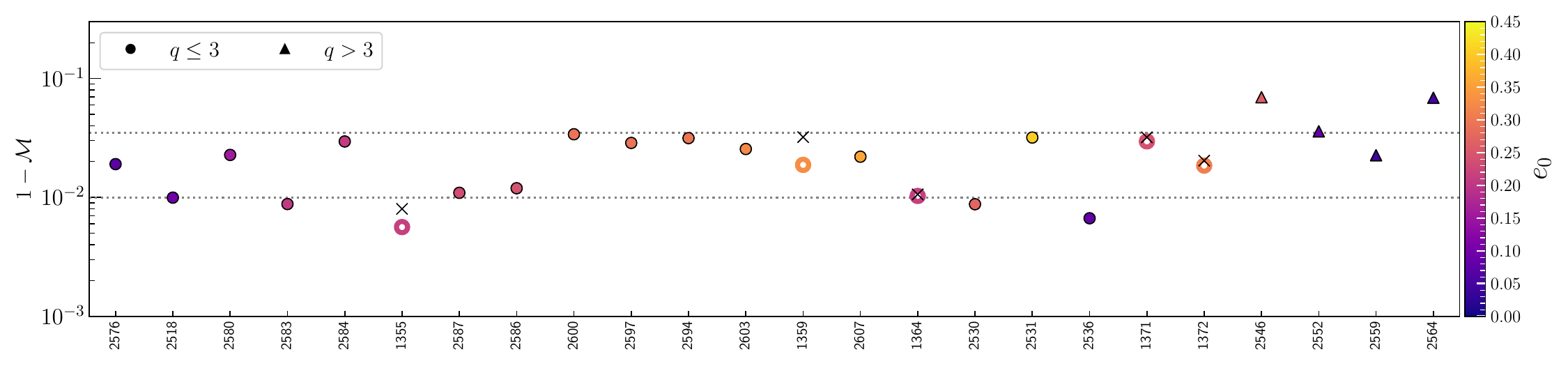}
    \includegraphics[width=\textwidth]{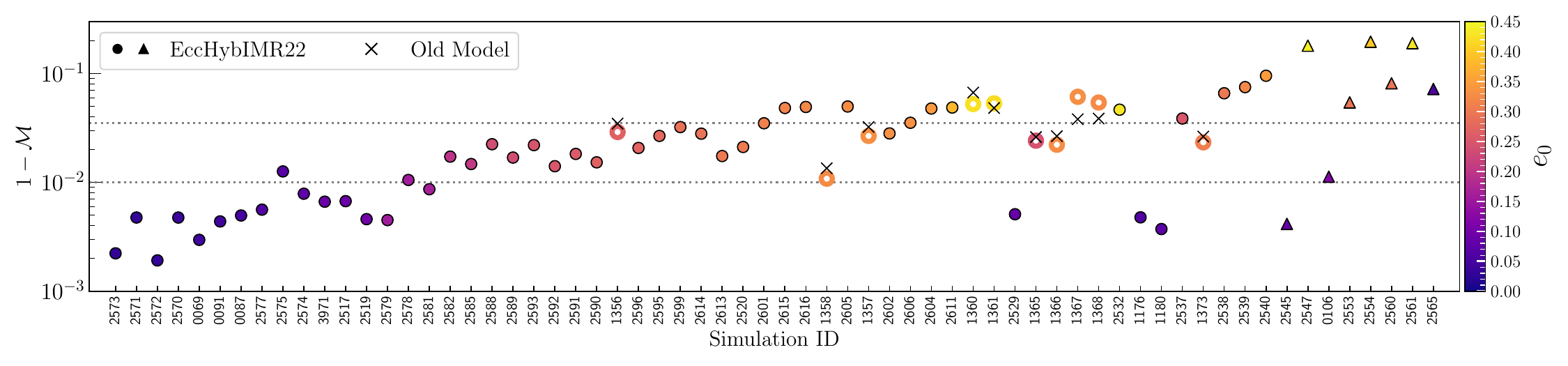}
    \caption{Mismatches ($1-\mathcal{M}$; see Eq.~\eqref{eq:match}) between our analytical model (\texttt{EccHybIMR22}) constructed in Sec.~\ref{sec:22-mode-model} and the \textit{testing set} (bottom panel) hybrids for a total mass of $40M_\odot$ at averaged 22 mode frequency of 20Hz using advanced LIGO sensitivity~\cite{TheLIGOScientific:2014jea}. For completeness, mismatch with 
    the \textit{training set} (top panel) hybrids are also shown. Properties of both sets of hybrids, \textit{training} and \textit{testing}, are listed in Table~\ref{tab:nrsims}. Filled circles (triangles) represent the hybrids corresponding to newly added simulations to the SXS catalog~\cite{Scheel:2025jct} with $q\leq3\ (q>3)$, whereas the hollow markers denote the old  ones~\cite{Hinder:2017sxy} used in constructing hybrids presented in Ref.~\cite{Manna:2024ycx}. The dotted lines denotes $96.5\%$ and $99\%$ matches respectively. The colorbar displays the initial eccentricity of the hybrid evaluated at 20Hz again assuming a total mass of 40$M_{\odot}$.}
    \label{scatter_comp_training_testing}
\end{figure*}

\section{Validation}
\label{sec:validation}
In this section we compare \texttt{EccHybIMR22} with testing set hybrids as well as against 175 NR simulations including several near circular simulations. Visual proof of its performance is summarised through Figs.~\ref{scatter_comp_training_testing}-\ref{ecc_hist} and in Figs.~\ref{strain_comp_training}-\ref{strain_comp_testing}. Figure~\ref{scatter_comp_training_testing} displays mismatches against testing set hybrids (\textit{bottom panel}) as well as against the set of hybrids used in training the model (\textit{top panel}) for completeness. These mismatches correspond to a fixed mass of $40M_\odot$ chosen arbitrarily to indicate representative mismatches. Note however, the mismatch typically drops (increases) as we move to heavier (lighter) binaries. Note also that, when comparing against \textit{testing set} hybrids the match is optimised by sampling over the initial parameters such as ($e_0, l_0, f_0$). 
We find that for $q\leq 3$, \texttt{EccHybIMR22} recovers most of the \textit{testing} set hybrids with matches $\gtrsim 96.5\%$ in the parameter ranges $0\lesssim e_0 \lesssim0.33;\ -\pi \leq l_0 \leq \pi$. However, for $q > 3$ systems, the hybrids are recovered with similar matches only for $e_0\lesssim 0.12$. This might be due to increased complexity of the underlying dynamics in the high eccentricity, high mass ratio regime. Similar observations were reported for comparisons of \texttt{IMRPhenomXE} against NR simulations in \cite{Ramos-Buades:2026kbq}.\\ 

For comparison, we also compute the match between the previous version of the model (referred to as `Old Model' in Fig.~\ref{scatter_comp_training_testing} and elsewhere in the paper) and the hybrids presented in Ref.~\cite{Manna:2024ycx} -- denoted by crosses in Fig.~\ref{scatter_comp_training_testing}. The mismatches between \texttt{EccHybIMR22} with these hybrids are shown as hollow markers. As can be seen in most cases \texttt{EccHybIMR22} performs better (even if marginal) compared to the old model. But more importantly, \texttt{EccHybIMR22} maintains its accuracy at low mass ratios ($q<3$) despite the extension of the mass ratio space (up to $q=10$) in this work.  
Finally, strain comparisons for each hybrid-model pair is displayed in Figs.~\ref{strain_comp_training}-\ref{strain_comp_testing} providing a visual proof of model's performance. Table~\ref{tab:nrsims} also explicitly lists the matches (last column) with cases highlighted as bold when the match drops below $96.5\%$.\\ 

Further, Fig.~\ref{fig:mismatch_NR} shows in a tile-plot the comparison against the NR simulations with $e_{\rm ref}^{\rm NR}\in[0,0.45]$ with the average mismatch computed within each bin.
Although, these comparisons provide an independent (and robust) check on our model performance, we can only probe masses higher than the values allowed by the length of each simulation. We estimate the total mass for each comparison using the advanced LIGO frequency cut-off of 20Hz and then compute the mismatch between the model and the NR simulation by again optimising over reference initial values --- ($e^{\rm NR}_{\rm ref}$, $l^{\rm NR}_{\rm ref}$) --- obtained from simulation metadata. We find that \texttt{EccHybIMR22} is able to recover ($140$ out of the $175$) target simulations with matches $\gtrsim96.5\%$ for systems with $0\lesssim e_{\rm ref}^{\rm NR}\lesssim0.3$. Further, the model is able to reproduce the circular limit (with mismatch $\lesssim1\%$) despite being trained on purely eccentric targets, suggesting a smooth recovery of circular templates. For comparison, we also show the mismatches with the old model; see left panel for Fig.~\ref{fig:mismatch_NR}. For $e_{\rm ref}^{\rm NR}\gtrsim0.1$, the new model exhibits a clear improvement (see middle panel of Fig.~\ref{fig:mismatch_NR}) in matches (up to $43\%)$ compared to the old model. Additionally, \texttt{EccHybIMR22} consistently provides higher match across $q>3$ parameter space; this is expected since the old model was only trained for mass ratios in the range $q\leq 3$. Specifically, \texttt{EccHybIMR22} (old model) provides matches $\gtrsim96.5\%$ against $\sim80\%\ (74\%)$ of the NR simulations used for validation. This enhanced accuracy is primarily driven by modelling the high $-\ q$ regime and likely also due to the use of increased number of simulations used to train the model. Note also that \texttt{EccHybIMR22} provides matches $\mathcal{M}\gtrsim96.5\%$ only for the NR simulations for which $e(t^{\mathcal{A,\omega}}_{\rm match})\lesssim0.15$ (as displayed in Fig.~\ref{ecc_hist}), where $t^{\mathcal{A,\omega}}_{\rm match}={\rm min} \{t^{\mathcal{A}}_{\rm match},\ t^\omega_{\rm match}\}$. This likely indicates that attaching a quasi-circular-merger-ringdown model may not be suitable choice beyond a certain point.\\

Finally, \texttt{EccHybIMR22} captures the dependence of $(t_{\rm shift},t_{\rm match}^{\mathcal{A}},t_{\rm match}^{\omega})$ on parameters ($q,e,l$) using basic Taylor or Fourier expansions. While this representation provides simple interpretable closed form expressions, its flexibility is limited for high non-linear parameter dependencies (in fact, we find that increasing the size of \textit{training set} results in poorer model fits.). To minimize the loss in the match while mapping from numerical to physical parameter space, a more flexible representation can be adopted using artificial neural network (ANN)~\cite{paszke2019pytorch}. We train an alternate model based on ANN using a \textit{training set} containing all hybrids providing $\gtrsim96.5\%$ match against the numerical model. To assess the performance of the model, we use it to recover the same set of NR simulations in right panel of Fig.~\ref{fig:mismatch_NR}. We find that, in the parameter space $e_{\rm ref}^{\rm NR}\gtrsim0.15$ and $q\leq8$, the ANN model is able to recover $\sim44\%$ of the simulations with $\mathcal{M}\gtrsim96.5\%$ as opposed to $\sim42\%$ using \texttt{EccHybIMR22}. However this improvement comes at the cost of reduced sensitivity in the high mass ratio regime; for example \texttt{EccHybIMR22} provides up to $2\%$ higher matches than the ANN model for systems with $q>8$; this is likely due to fewer hybrids at high mass ratios. Overall, the ANN model recovers $\sim81\%$ of the total NR simulations (as opposed to $\sim80\%$ with \texttt{EccHybIMR22}) with $\mathcal{M}\gtrsim96.5\%$.

\begin{figure*}[htbp!]
    \centering
    \includegraphics[width=\linewidth]{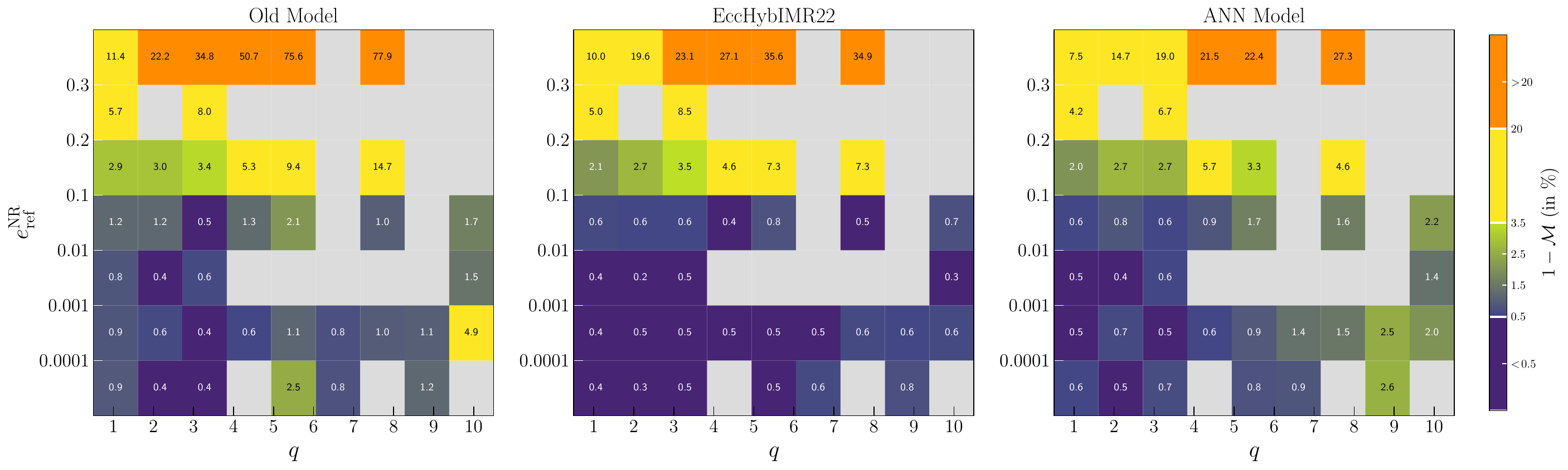}
    \caption{Mismatches between \texttt{EccHybIMR22} and 175 NR simulations (including circular simulations) as functions of mass-ratio ($q$) and reference NR eccentricity ($e_{\rm ref}^{\rm NR})$. The left, middle and right panels show the results for the `Old Model'~\cite{Manna:2024ycx}, \texttt{EccHybIMR22} and an artificial neural network (ANN) model (see Sec.~\ref{sec:validation}), respectively. The average mismatch (in $\%$) computed within each bin is reported by the colorbar. The mismatches are computed for the lowest total mass (of the respective NR simulation) that can enter the advanced LIGO band at 20Hz. The old model, \texttt{EccHybIMR22} and ANN model recover $\sim 74\%,\ 80\%,$ and $81\%$ of the NR simulations with matches $\gtrsim96.5\%$ respectively.}
    
    \label{fig:mismatch_NR}
\end{figure*}

\begin{figure}[htbp!]
    \includegraphics[width=0.49
    \textwidth]{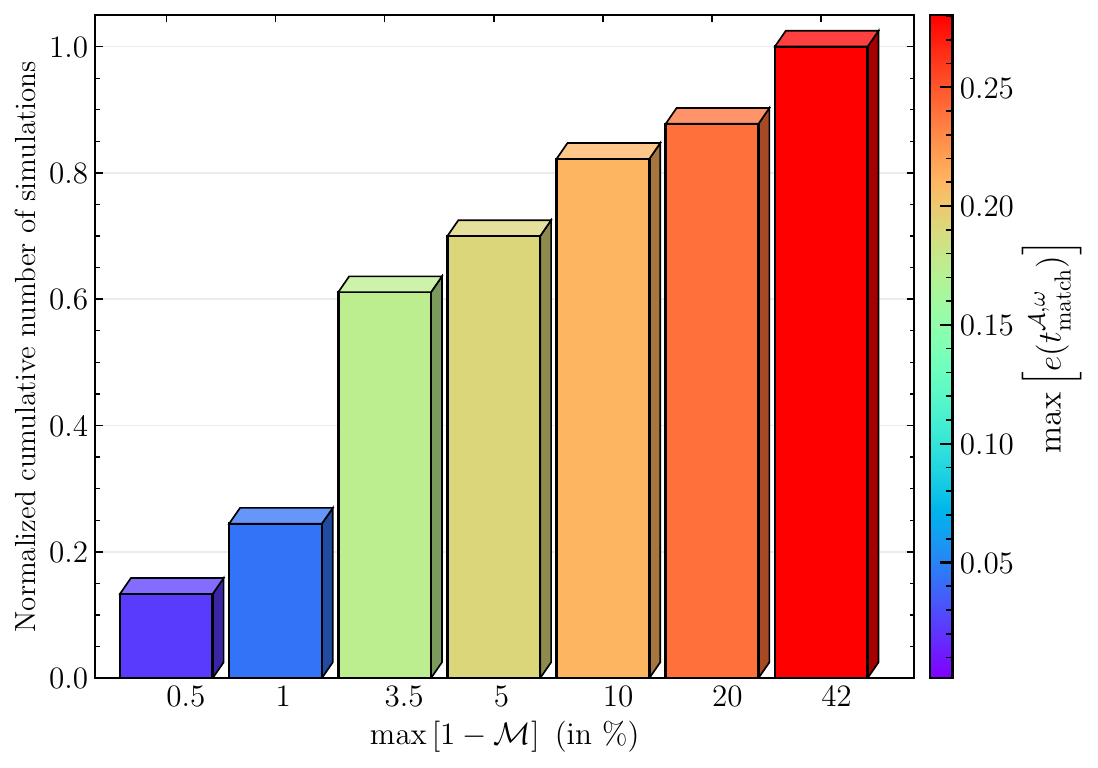}
    \caption{Cumulative histogram showing the fraction of NR simulation utilized in constructing the hybrids as a function of maximum mismatch corresponding to each bin. The colorbar reports the maximum of eccentricity ($e$) at the attachment time $t^{\mathcal{A},\omega}_{\rm match}$ of the simulations contained within each bin, where $t^{\mathcal{A,\omega}}_{\rm match}$ is defined as ${\rm min} \{t^{\mathcal{A}}_{\rm match},\ t^\omega_{\rm match}\}$. Simulations with $e(t^{\mathcal{A,\omega}}_{\rm match})\lesssim0.15$ are recovered by \texttt{EccHybIMR22} with $\mathcal{M}\gtrsim96.5\%$.} 
\label{ecc_hist}
\end{figure}

\section{Summary and Conclusions}
\label{sec:disc}
This article aims to construct a hybrid model obtained by attaching an eccentric inspiral to a quasi-circular-merger-ringdown model suitable for binary black hole mergers on eccentric orbits. We refer this model throughout this paper as \texttt{EccHybIMR22}. Specifically, leveraging the recent update to SXS catalog of NR simulations \cite{Scheel:2025jct}, we extend the 22 mode model presented in Refs.~\cite{Chattaraj:2022tay,Manna:2024ycx} to large mass ratios --- $3<q\leq 10$. In order to be sensitive to low total mass systems, we first construct a set of IMR hybrids by combining waveforms from the PN approach with NR simulations when the agreement between the two in a certain window is better than 96\% (see for instance Fig.~\ref{fig:PN-NR_scatter}). These hybrids are subsequently categorised into two sets --- \textit{training and testing} --- and broadly represent the parameter space that we intend to explore here. Figure~\ref{fig:hybridplot} shows a set of hybrids for seven different $q$ values. These hybrids are used as targets for both calibration and validation purposes. First, a numerical model is obtained by smooth attachment of the inspiral and merger-ringdown piece and compared against training set hybrids. Analytical fits for time-shits and attachment times are subsequently obtained which then are used to generate \texttt{EccHybIMR22}. (See Sec.~\ref{sec:22-mode-model} and Appendix~\ref{sec:appn_fits} for details.) The model is then validated against testing set hybrids as well as against NR simulations including several near circular simulations to validate the model in circular limit; see Sec.~\ref{sec:validation} for details. Finally, we provide an alternate model based on ANN, which provides more flexibility over simplistic Taylor/Fourier expansion based functional dependence used in the analytical fits and minimizes the loss in $\mathcal{M}$ while mapping from numerical to arbitrary parameter space. The ANN model though performs slightly better for $q<8$ and high eccentric cases compared to \texttt{EccHybIMR22}, the performance of the latter is better at larger mass ratios (see middle and right panels of Fig.~\ref{fig:mismatch_NR}). This is likely due to fewer NR simulations at high mass ratios. Note that the ANN model utilizes all hybrids providing $96.5\%$ match against the numerical model for training and not just the \textit{training set}.\\ 

When comparing the performance of a recent frequency-domain eccentric model \texttt{IMRPhenomXE}~\cite{Ramos-Buades:2026kbq} against NR simulations, we find that our time-domain \texttt{EccHybIMR22} model recovers a comparable fraction of NR simulations ($\sim77\%$) with mismatches $\lesssim3\%$, as opposed to $\sim72\%$ with \texttt{IMRPhenomXE} for the respective target simulation set.\footnote{Note that, Ref.~\cite{Ramos-Buades:2026kbq} chooses to set $f_{\rm min}$ to $1.35\times\tilde{f}_{\rm start}$, where $\tilde{f}_{\rm start}$ is the initial orbit-averaged frequency of the respective NR simulation, when the target NR simulation does not span the frequency range of (10, 2048)Hz and reports maximum mismatch ($\mathcal{M}_{\rm max}$) for systems with total mass in the range (20, 200)$M_{\odot}$. While in our case, the match is computed assuming a low frequency cut-off of 20Hz as the initial start frequency for all simulations.} Note also that,  \texttt{IMRPhenomXE} is validated against a slightly larger set of simulations (186) than our analysis (175) and thus our recovery fraction is effectively larger (even though marginal) than the one quoted above. Nevertheless, \texttt{IMRPhenomXE} seems to have better sensitivity to large eccentricity and large mass ratios compared to ours. This might be due to their use of a better inspiral model compared to the one utilized here. For instance, the orbital phase is obtained by solving equations governing the binary dynamics at 3PN order compared to 2PN in \texttt{EccentricTD} and mode amplitudes are $\mathcal{O}(e^{12})$ accurate compared to $\mathcal{O}(e^6)$ here where $\mathcal{O}(e^n)$ indicates that corrections due to eccentricity beyond $n^{\rm th}$ are ignored. Further, inclusion of spins into our model would again require using an alternate inspiral model since \texttt{EccentricTD} is a nonspinning model.  Further, as demonstrated in Fig.~\ref{ecc_hist} sub-threshold mismatches ($<96.5\%$) with NR simulations with $e^{\rm NR}_{\rm ref}\gtrsim0.15$ hints possibly at the inadequacy of the use of QC model for the merger-ringdown portion of the model. Finally, \texttt{EccHybIMR22} is only a quadrupole mode model while exploring mass asymmetries necessarily requires modelling non-quadrupole harmonics as well; for instance a number of modes (odd $\ell+m$ modes) only contribute for unequal mass systems. While extensions including the effect of higher modes can be obtained following the proposal of our earlier works~\cite{Chattaraj:2022tay, Manna:2024ycx} or other proposals such as~\cite{Islam:2026blk}, modelling of higher order modes together with matters outlined above will be taken up in future extensions of our model.\\

For systems with total masses $\gtrsim30M_\odot$, the waveform generation time of \texttt{EccHybIMR22} typically lies within $\sim3$ seconds for mass ratio $q\leq10$ as shown in Fig.~\ref{fig:computational time}. However, for lower masses (for example $10M_\odot$) the computational time can go up to $\sim27$ seconds (See Appendix~\ref{sec: appn_cost} for more details). \texttt{EccHybIMR22} can be accessed publicly via Ref.~\cite{githubRepo}.

\section{Acknowledgments} 
We thank Kaushik Paul for useful comments and suggestions on our manuscript. We are grateful to the SXS Collaboration for releasing the updated catalog of numerical relativity simulations. Computations were performed in the CIT cluster provided by the LIGO laboratory. P.M. thanks Surajit Kalita for his guidance on artifical neural network methods. T.R.C. acknowledges the support of the National Science Foundation award PHY-2513124. C.K.M. acknowledges the support of SERB's Core Research Grant No.~CRG/2022/007959.\\

This document has LIGO preprint number LIGO-P2600540. 

\appendix
\section{Table of hybrids}
\label{sec:appn_hybrids}
The details of all target hybrids constructed in this work are provided in Table~\ref{tab:nrsims}. The hybrids are arranged by ascending $q$ and then by ascending $e_0$. The hybrids are split into \textit{training} and \textit{testing sets} determined based on mismatch between the numerical model and the hybrids as described in Sec.~\ref{sec:22-mode-model}. For reference, the matches ($\mathcal{M}$ in $\%$) with \texttt{EccHybIMR22} (presented in Sec.~\ref{sec:22-mode-model}) are also listed. The table lists a total of 91 hybrids, 24 of them are \textit{training} hybrids; the rest belong to \textit{testing} set. Three hybrids indicated in bold are not used for validation as their parameters lie sufficiently outside the calibration range. 

\section{Time domain model fits}
\label{sec:appn_fits}
We optimize the numerical model against the hybrids in the parameter ranges $-400M\leq t_{\rm shift}\leq400M,\ -1000M \leq t_{\rm match}^{\mathcal{A}}\leq -50M,\ -8000M \leq t_{\rm match}^{\omega} \leq -500M$. Next, using \textit{training} set hybrids, the analytical expressions for these parameters are constructed as shown below.
\vspace{-0.5em}
{
\begin{align}
t_{\rm shift}(q,e,l)
&=
A_{k\alpha\beta}\,
e^{\alpha}\,
q^{\beta}\,
\cos\!\left[
\left(\gamma_k + \delta_k e\right)l
+ a_k
\right]
\label{eq:t_shift_analytical}
\end{align}
\vspace{-1.5em}
\begin{align}
\label{eq:t_match_amplitude}
t^\mathcal{A}_{\rm match}(q,e,l)
=
\left[
A_{0\alpha\beta}
+
A_{k\alpha\beta}\cos(k l + a_k)
\right]
e^\alpha q^\beta
\end{align}
\vspace{-1.5em}
\begin{align}
    \label{eq:tmatch_frequency}
    t^\omega_{\rm match}(q,e,l)
    =
    \left[
    A_{0\alpha\beta}
    +
    A_{k\alpha\beta}\cos(k l)
    \right]
    e^\alpha q^\beta
\end{align}

The repeated indices are summed over in these expressions. The coefficients of the analytical fits are listed in Tables~\ref{tab:tshift-table} - \ref{tab:tfreq-table}. We obtain average relative errors of $\sim4\%$, $9\%$ and $3\%$ between the numerical and analytical estimates for $t_{\rm shift},\ t^\mathcal{A}_{\rm match}$, and $t^\omega_{\rm match}$, respectively.
}

\begin{figure}[htbp!]
    \includegraphics[width=0.49
    \textwidth]{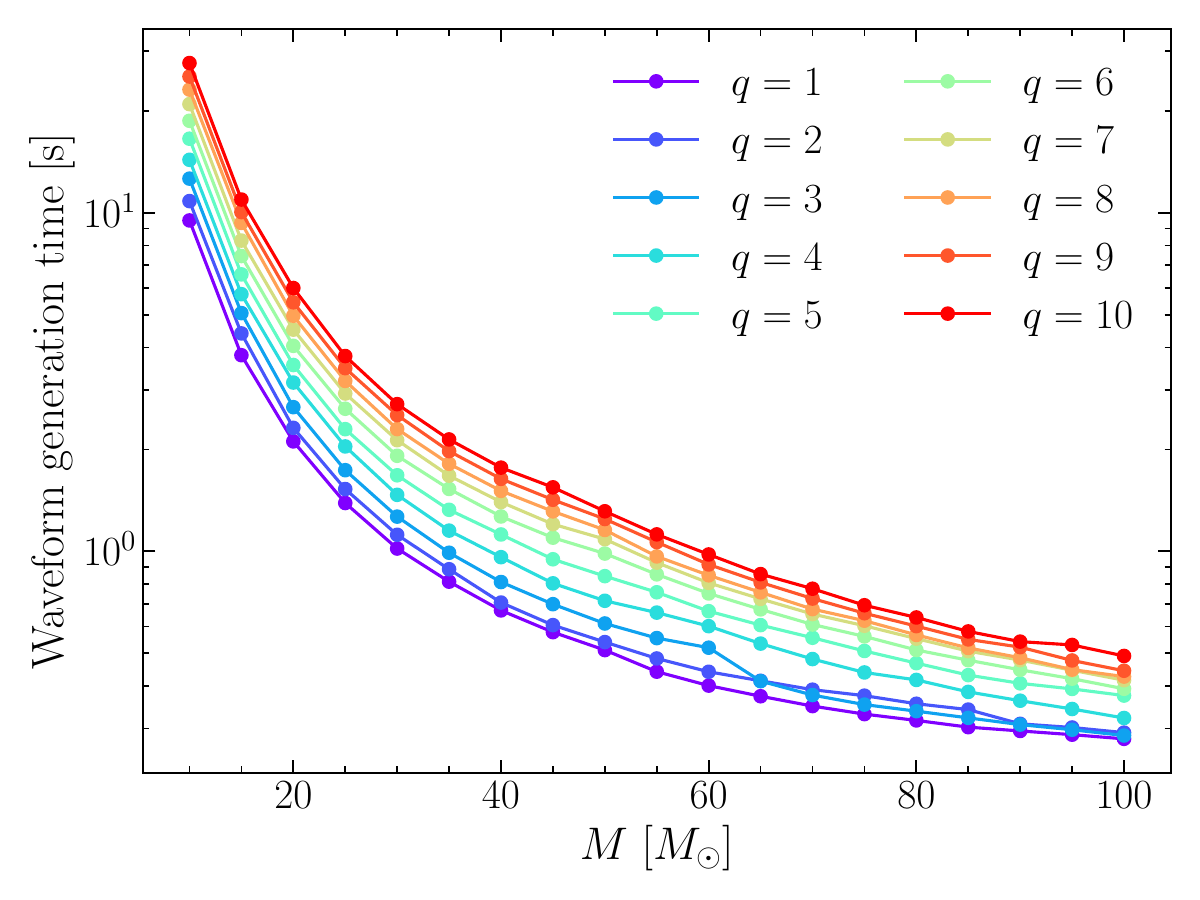}
    \caption{Average evaluation cost of \texttt{EccHybIMR22} sampled at 8192 Hz as a function of total mass for distinct mass ratios in the range $q\in[1,10]$ using an initial frequency of 10Hz and initial eccentricity $~0.3$. For $M\gtrsim30M_\odot$, waveform generation time typically lies within $\sim 3$ seconds. However, for lower total masses, the evaluation time goes up to $\sim27$ seconds.} 
    \label{fig:computational time}
\end{figure}

\section{Comparisons of GW Strain}
\label{sec:appn_strain}

To visualize the performance of \texttt{EccHybIMR22}, we provide strain ($h_+$) comparison plots of the model against the \textit{training} and \textit{testing set} hybrids in Fig.~\ref{strain_comp_training} and Fig.~\ref{strain_comp_testing}, respectively. Each plot shows \texttt{EccHybIMR22} plotted on top of the hybrid while the inset shows the same close to the merger, and we see general agreement between the two indicating the quality of the model.

\section{Computational cost}
\label{sec: appn_cost}

In this section, we evaluate the computation cost of \texttt{EccHybIMR22} in the total mass range $M\in[10,100]M_\odot$. We average over 10000 configurations for distinct mass ratios in the range $q\in[1,10]$. Assuming an initial frequency of 10Hz, initial eccentricity $\sim0.3$ and sampling rate of 8192 Hz, the average waveform generation time typically lies within $\sim 3$ seconds for systems with $M\gtrsim30M_\odot$ (see Fig.~\ref{fig:computational time}). Although, for lower masses (for example, $10M_\odot$), the evaluation time can reach up to $\sim27$ seconds. Note also that, for a fixed $M$, the average evaluation time increases with $q$.   

\begin{table*}[htbp!]
\setlength{\tabcolsep}{8pt} 
\caption{
Time-domain target hybrids constructed by matching NR simulations from the updated SXS Catalog and state-of-the-art PN prescriptions within a $1000M$ time window with initial time $t_{\rm w}$. SXS IDs are retained for identification against the BBH NR simulations used in constructing the hybrids. Each hybrid starts at fixed initial frequency $x_0=0.0337$ with parameters ($e_0,l_0$) such that at a reference frequency ($x^{\rm gw}_{\rm ref}$) it attains the parameters ($e^{\rm gw}_{\rm ref}, l^{\rm gw}_{\rm ref}$) obtained using the {\fontfamily{qcr}\selectfont{gw\_eccentricity}} package based on Refs.~\cite{Shaikh:2023ypz,Shaikh:2025tae}.
The PN frequency parameter ($x$) is defined as $x_{\rm gw}$ =$(\pi M f_{\rm gw})^{2/3}$ where $f_{\rm gw}$ and $M$ denote the GW frequency of the dominant mode and the total mass of the binary. We also report the mass ratio ($q$), number of orbital cycles ($N_{\rm orb}$), initial time of hybridization window ($t_{\rm w}$) and match ($\mathcal{M}$) between the hybrids and the new model (presented in Sec.~\ref{sec:22-mode-model}). The hybrids are first arranged by ascending $q$ and then by ascending $e_0$. Hybrids marked in bold font are sufficiently outside the calibration range of the model and hence are not used for validation. Additionally, $\mathcal{M}<96.5\%$ cases are also indicated in bold.
}
\vspace{0.2cm}

\begin{tabular}{clcccccrrrc}
  \hline
  \hline
Count & Simulation ID & $q$ & $x^{\rm gw}_{\rm ref}$ & $e^{\rm gw}_{\rm ref}$ & $l^{\rm gw}_{\rm ref}$ & $e_0$ & $l_0$ & $N_{\rm orb}$ & $t_{\rm w}$ & $\mathcal{M}\ (\%)$ \\
\addlinespace[0.1cm]
\hline
\multicolumn{11}{c}{\textbf{Training Set}}\\
1 & HYB:SXS:BBH:2576 & 1 & 0.0573 & 0.030 & 1.112 & 0.073 & -1.788 & 91.5 & -6000 & 98.1\\
2 & HYB:SXS:BBH:2518 & 1 & 0.0468 & 0.056 & 4.367 & 0.096 & -2.318 & 89.9 & -12000 & 99.0\\
3 & HYB:SXS:BBH:2580 & 1 & 0.0557 & 0.068 & 3.782 & 0.155 & -0.509 & 86.2 & -6000 & 97.7\\
4 & HYB:SXS:BBH:2583 & 1 & 0.0575 & 0.088 & 3.142 & 0.205 & -2.162 & 80.8 & -6000 & 99.1\\
5 & HYB:SXS:BBH:2584 & 1 & 0.0566 & 0.091 & 2.684 & 0.207 & -1.098 & 80.3 & -4000 & 97.0\\
6 & HYB:SXS:BBH:1355 & 1 & 0.0389 & 0.173 & 2.455 & 0.214 & -0.360 & 79.5 & -2000 & 99.4\\
7 & HYB:SXS:BBH:2587 & 1 & 0.0527 & 0.119 & 3.401 & 0.235 & 1.988 & 76.6  & -8000 & 98.9\\
8 & HYB:SXS:BBH:2586 & 1 & 0.0574 & 0.109 & 3.770 & 0.246 & -0.655 & 76.0 & -4000 & 98.8\\
9 & HYB:SXS:BBH:2597 & 1 & 0.0531 & 0.153 & 2.133 & 0.293 & 0.862 &  69.7 & -4000 & 97.1\\
10 & HYB:SXS:BBH:2600 & 1 & 0.0479 & 0.180 & 3.086 & 0.293 & 0.965 & 69.6 & -4000 & 96.6\\
11 & HYB:SXS:BBH:2594 & 1 & 0.0553 & 0.146 & 2.793 & 0.297 & 2.181 & 68.4 & -6000 & 96.9\\

12 & HYB:SXS:BBH:2603 & 1 & 0.0459 & 0.216 & 2.318 & 0.323 & -0.198 & 64.7 & -2000 & 97.4\\
13 & HYB:SXS:BBH:1359 & 1 & 0.0347 & 0.317 & 1.131 & 0.328 & 0.680 & 63.1 & -2000 & 98.1\\
14 & HYB:SXS:BBH:2607 & 1 & 0.0473 & 0.236 & 4.171 & 0.359 & 0.967 & 58.4 & -6000 & 97.8\\
15 & HYB:SXS:BBH:1364 & 2 & 0.0391 & 0.172 & 2.681 & 0.214 & -1.144 & 89.3 & -2000 & 99.0\\
16 & HYB:SXS:BBH:2530 & 2 & 0.0522 & 0.144 & 2.479 & 0.272 & 1.713 & 81.0 & -6000 & 99.1\\
17 & HYB:SXS:BBH:2531 & 2 & 0.0528 & 0.238 & 4.148 & 0.405 & 1.757 & 57.1 & -2000 & 96.8\\
18 & HYB:SXS:BBH:2536 & 3 & 0.0453 & 0.049 & 2.724 & 0.081 & -3.001 & 120.0 & -12000 & 99.3\\
19 & HYB:SXS:BBH:1371 & 3 & 0.0380 & 0.204 & 2.621 & 0.242 & -1.228 & 99.8 & -2000 & 97.0\\
20 & HYB:SXS:BBH:1372 & 3 & 0.0344 & 0.300 & 1.789 & 0.308 & -1.962 & 88.0 & -2000 & 98.2\\
21 & HYB:SXS:BBH:2546 & 4 & 0.0503 & 0.146 & 2.805 & 0.263 & -2.336 & 111.8 & -10000 & \textbf{93.0}\\
22 & HYB:SXS:BBH:2552 & 6 & 0.0562 & 0.034 & 2.547 & 0.081 & 0.861 & 181.4 & -14000 & \textbf{96.4}\\
23 & HYB:SXS:BBH:2559 & 8 & 0.0673 & 0.099 & 1.822 & 0.041 & -1.767 & 226.9 & -8000 & 97.7\\
24 & HYB:SXS:BBH:2564 & 10 & 0.0694 & 0.014 & 2.142 & 0.050 & 0.070 & 269.8 & -8000 & \textbf{93.1}\\
\multicolumn{11}{c}{\textbf{Testing Set}}\\
25 & HYB:SXS:BBH:2573 & 1 & 0.0557 & 0.013 &  3.590 & 0.031 & -2.806 & 93.1 & -6000 & 99.8\\
26 & HYB:SXS:BBH:2571 & 1 & 0.0564 & 0.014 &  2.132 & 0.033 & 0.003 & 92.8 & -6000 & 99.5\\
27 & HYB:SXS:BBH:2572 & 1 & 0.0565 & 0.014 &  2.581 & 0.033 & 1.046 & 92.7 & -6000 & 99.8\\
28 & HYB:SXS:BBH:2570 & 1 & 0.0586 & 0.015 & 2.584 & 0.039 & 0.056 & 92.7 & -6000 & 99.5\\
29 & HYB:SXS:BBH:0069 & 1 & 0.0568 & 0.018 & 5.958 & 0.044 & 0.073 & 92.8 & -6000 & 99.7\\
30 & HYB:SXS:BBH:0091 & 1 & 0.0545 & 0.020 & 5.665 & 0.046 & -2.392 & 92.7 & -8000 & 99.6\\
31 & HYB:SXS:BBH:0087 & 1 & 0.0581 & 0.020 & 0.673 & 0.050 & 2.110 & 92.9 & -6000 & 99.5\\
32 & HYB:SXS:BBH:2577 & 1 & 0.0559 & 0.027 & 1.835 & 0.064 & -3.047 & 91.8 & -4000 & 99.4\\
33 & HYB:SXS:BBH:2575 & 1 & 0.0586 & 0.026 & 1.956 & 0.067 & -0.373 & 91.6 & -6000 & 98.7\\
34 & HYB:SXS:BBH:2574 & 1 & 0.0548 & 0.034 & 3.551 & 0.077 & -1.841 & 90.9 & -6000 & 99.2\\
35 & HYB:SXS:BBH:3971 & 1 & 0.0646 & 0.030 & 2.139 & 0.091 & 0.221 & 90.3 & -4000 & 99.3\\
36 & HYB:SXS:BBH:2517 & 1 & 0.0646 & 0.030 & 2.139 & 0.092 & 0.224 & 90.4 & -4000 & 99.3\\
37 & HYB:SXS:BBH:2519 & 1 & 0.0520 & 0.047 & 3.538 & 0.097 & -2.875 & 89.9 & -10000 & 99.5\\
38 & HYB:SXS:BBH:2579 & 1 & 0.0534 & 0.073 & 3.503 & 0.155 & 3.118 & 85.6 & -6000 & 99.5\\
39 & HYB:SXS:BBH:2578 & 1 & 0.0532 & 0.075 & 4.062 & 0.158 & 2.377 & 85.9 & -6000 & 98.9\\
40 & HYB:SXS:BBH:2581 & 1 & 0.0550 & 0.073 & 2.285 & 0.162 & 0.199 & 85.7 & -6000 & 99.1\\
41 & HYB:SXS:BBH:2582 & 1 & 0.0537 & 0.094 & 3.170 & 0.196 & 0.260 & 81.3 & -6000 & 98.3\\
42 & HYB:SXS:BBH:2585 & 1 & 0.0522 & 0.103 & 2.502 & 0.204 & 2.034 & 80.9 & -6000 & 98.5\\
43 & HYB:SXS:BBH:2588 & 1 & 0.0506 & 0.127 & 3.336 & 0.235 & -1.091 & 77.1 & -4000 & 97.8\\
44 & HYB:SXS:BBH:2589 & 1 & 0.0526 & 0.121 & 3.459 & 0.238 & 1.518 & 76.7 & -6000 & 98.3\\

\addlinespace[0.25cm]
\hline
\hline

\bottomrule
\end{tabular}
\label{tab:nrsims}
\end{table*}

\addtocounter{table}{-1}
\begin{table*}[ht!]
\setlength{\tabcolsep}{8pt}
\caption{\textit{Continued.}}
\vspace{0.2cm}

\begin{tabular}{clcccccrrrc}
  \hline
  \hline
Count & Simulation ID & $q$ & $x^{\rm gw}_{\rm ref}$ & $e^{\rm gw}_{\rm ref}$ & $l^{\rm gw}_{\rm ref}$ & $e_0$ & $l_0$ & $N_{\rm orb}$ & $t_{\rm w}$ & $\mathcal{M}\ (\%)$\\
\addlinespace[0.1cm]
\hline

45 & HYB:SXS:BBH:2593 & 1 & 0.0537 & 0.122 & 3.142 & 0.247 & 2.772 & 75.3 & -6000 & 97.8\\
46 & HYB:SXS:BBH:2592 & 1 & 0.0563 & 0.115 & 3.267 & 0.251 & -0.446 & 74.8 & -4000 & 98.6\\
47 & HYB:SXS:BBH:2591 & 1 & 0.0571 & 0.115 & 2.364 & 0.255 & 3.053 & 75.2 & -6000 & 98.2\\
48 & HYB:SXS:BBH:2590 & 1 & 0.0557 & 0.126 & 4.528 & 0.266 & -1.852 & 72.8 & -6000 & 98.5\\
49 & HYB:SXS:BBH:1356 & 1 & 0.0375 & 0.230 & 1.717 & 0.266 & 1.601 & 73.0 & -2000 & 97.1\\
50 & HYB:SXS:BBH:2596 & 1 & 0.0503 & 0.152 & 2.970 & 0.272 & -0.891 & 72.3 & -6000 & 97.9\\
51 & HYB:SXS:BBH:2595 & 1 & 0.0514 & 0.150 & 3.632 & 0.276 & 2.247 & 71.9 &-6000 & 97.3\\
52 & HYB:SXS:BBH:2599 & 1 & 0.0495 & 0.169 & 2.524 & 0.290 & 0.726 & 69.5 & -6000 & 96.8\\
53 & HYB:SXS:BBH:2614 & 1 & 0.0510 & 0.163 & 3.288 & 0.292 & 0.436 & 68.5 & -6000 & 97.2\\
54 & HYB:SXS:BBH:2613 & 1 & 0.0508 & 0.169 & 2.793 & 0.300 & -0.824 & 68.7 & -6000 & 98.3\\
55 & HYB:SXS:BBH:2520 & 1 & 0.0671 & 0.109 & 2.715 & 0.304 & -1.253 &  68.0 & -2000 & 97.9\\
56 & HYB:SXS:BBH:2601 & 1 & 0.0559 & 0.151 & 3.239 & 0.310 & 0.457 & 66.3 & -6000 & 96.5\\
57 & HYB:SXS:BBH:2615 & 1 & 0.0467 & 0.206 & 2.786 & 0.317 & 0.116 & 65.0 & -2000 & \textbf{95.2}\\
58 & HYB:SXS:BBH:2616 & 1 & 0.0496 & 0.191 & 3.491 & 0.321 & -0.361 & 65.2 & -6000 & \textbf{95.1}\\
59 & HYB:SXS:BBH:1358 & 1 & 0.0340 & 0.322 & 1.215 & 0.325 & -1.878 & 64.5 & -2000 & 98.9\\
60 & HYB:SXS:BBH:2605 & 1 & 0.0495 & 0.196 & 2.709 & 0.327 & -1.165 & 63.8 & -4000 & \textbf{95.0}\\
61 & HYB:SXS:BBH:1357 & 1 & 0.0344 & 0.322 & 1.215 & 0.330 & -0.048 & 63.8 & -2000 & 97.3\\
62 & HYB:SXS:BBH:2602 & 1 & 0.0525 & 0.184 & 3.746 & 0.333 & 1.946 & 63.8 & -6000 & 97.2\\
63 & HYB:SXS:BBH:2606 & 1 & 0.0471 & 0.216 & 3.770 & 0.333 & 0.866 & 63.5 & -2000 & 96.5\\
64 & HYB:SXS:BBH:2604 & 1 & 0.0502 & 0.204 & 4.262 & 0.343 & 1.001 & 61.4 & -6000 & \textbf{95.2}\\
65 & HYB:SXS:BBH:2611 & 1 & 0.0552 & 0.204 & 4.168 & 0.380 & 2.174 & 55.3 & -4000 & \textbf{95.1}\\
66 & HYB:SXS:BBH:1361 & 1 & 0.0644 & 0.187 & 3.691 & 0.420 & 1.130 & 48.7 & -2000 & \textbf{94.7}\\
67 & HYB:SXS:BBH:1360 & 1 & 0.0648 & 0.187 & 3.691 & 0.424 & 2.573 & 48.8 & -2000 & \textbf{94.8}\\
68 & \textbf{HYB:SXS:BBH:2521} & 1 & 0.0526 & 0.254 & 1.879 & 0.442 & 3.089 & 57.6 & -6000 & -\\
69 & \textbf{HYB:SXS:BBH:1362} & 1 & 0.0671 & 0.223 & 3.396 & 0.483 & -2.761 & 38.4 & -2000 & -\\
70 & HYB:SXS:BBH:2529 & 2 & 0.0593 & 0.033 & 3.416 & 0.086 & -2.164 & 101.2 & -6000 & 99.5\\
71 & HYB:SXS:BBH:1365 & 2 & 0.0376 & 0.209 & 2.262 & 0.243 & -0.987 & 84.9 & -2000 & 97.6\\
72 & HYB:SXS:BBH:1368 & 2 & 0.0338 & 0.324 & 1.382 & 0.325 & -3.108 & 71.4 & -2000 & \textbf{94.6}\\
73 & HYB:SXS:BBH:1366 & 2 & 0.0344 & 0.320 & 1.299 & 0.328 & -0.657 & 70.8 & -2000 & 97.8\\
74 & HYB:SXS:BBH:1367 & 2 & 0.0346 & 0.320 & 1.299 & 0.329 & -2.575 & 70.4 & -2000 & \textbf{93.9}\\
75 & HYB:SXS:BBH:2532 & 2 & 0.0610 & 0.212 & 2.912 & 0.431 & -2.310 & 52.0 & -2000 & \textbf{95.3}\\
76 & \textbf{HYB:SXS:BBH:1369} & 2 & 0.0605 & 0.272 & 2.451 & 0.498 & -0.942 & 40.3 & -2000 & -\\
77 & HYB:SXS:BBH:1176 & 3 & 0.0789 & 0.014 & 2.315 & 0.063 & 0.992 & 121.0 & -2000 & 99.5\\
78 & HYB:SXS:BBH:1180 & 3 & 0.0767 & 0.017 & 3.383 & 0.073 & -2.608 & 120.4 & -2000 & 99.6\\
79 & HYB:SXS:BBH:2537 & 3 & 0.0531 & 0.126 & 2.760 & 0.250 & 2.164 & 98.7 & -4000 & \textbf{96.1}\\
80 & HYB:SXS:BBH:2538 & 3 & 0.0390 & 0.249 & 2.223 & 0.300 & -2.519 & 89.4 & -2000 & \textbf{93.4}
\\
81 & HYB:SXS:BBH:1373 & 3 & 0.0344 & 0.300 & 1.789 & 0.308 & -1.962 & 87.6 & -2000 & 97.7\\
82 & HYB:SXS:BBH:2539 & 3 & 0.0499 & 0.189 & 2.356 & 0.321 & -0.678 & 84.7 & -6000 & \textbf{92.5}\\
83 & HYB:SXS:BBH:2540 & 3 & 0.0636 & 0.143 & 2.557 & 0.346 & 2.813 & 80.1 & -2000 & \textbf{90.4}\\
84 & HYB:SXS:BBH:2545 & 4 & 0.0580 & 0.034 & 4.937 & 0.086 & 2.692 & 138.7 & -8000 & 99.6\\
85 & HYB:SXS:BBH:2547 & 4 & 0.0470 & 0.308 & 1.512 & 0.437 & 0.341 & 70.1 & -6000 & \textbf{81.9}\\
86 & HYB:SXS:BBH:0106 & 5 & 0.0738 & 0.031 & 2.452 & 0.121 & 2.051 & 156.0 & -4000 & 98.9\\
87 & HYB:SXS:BBH:2553 & 6 & 0.0591 & 0.128 & 2.011 & 0.293 & -2.585 & 136.5 & -10000 & \textbf{94.6}\\
88 & HYB:SXS:BBH:2554 & 6 & 0.0535 & 0.228 & 2.475 & 0.398 & -0.118 & 102.1 & -14000 & \textbf{80.4}\\
89 & HYB:SXS:BBH:2560 & 8 & 0.0549 & 0.145 & 1.868 & 0.294 & 2.970 & 167.9 & -14000 & \textbf{91.9}\\
90 & HYB:SXS:BBH:2561 & 8 & 0.0570 & 0.232 & 4.398 & 0.430 & 1.565 & 113.9 & -10000 & \textbf{81.0}\\
91 & HYB:SXS:BBH:2565 & 10 & 0.0694 & 0.017 & 2.189 & 0.060 & 0.192 & 268.9 & -8000 & \textbf{92.8}\\

\addlinespace[0.25cm]
\hline
\hline
\end{tabular}
\label{tab:nrsims}
\end{table*}

\begin{figure*} 
\centering
    \includegraphics[width=0.245\textwidth]{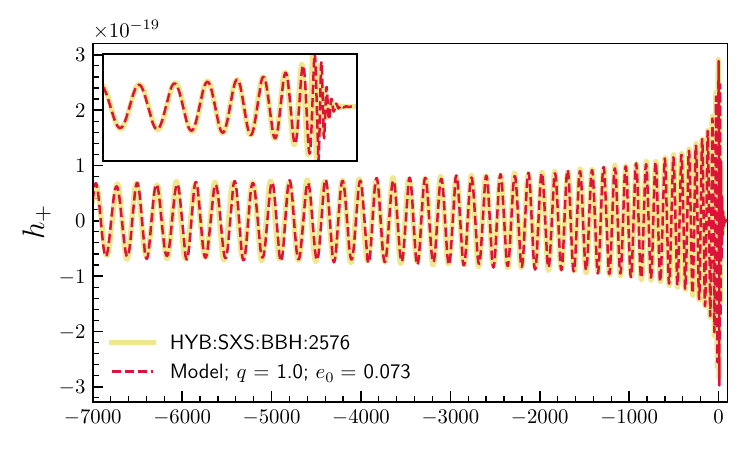}
    \includegraphics[width=0.245\textwidth]{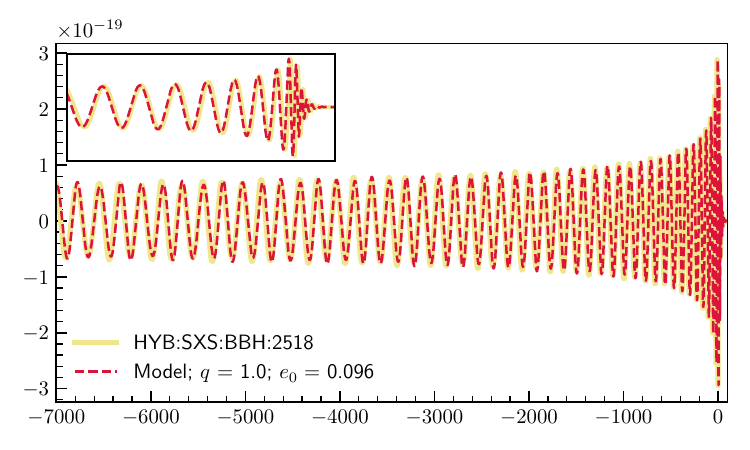}
    \includegraphics[width=0.245\textwidth]{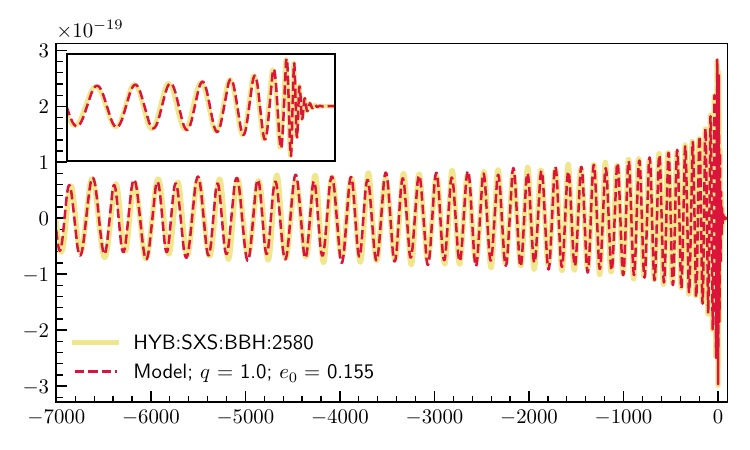}
    \includegraphics[width=0.245\textwidth]{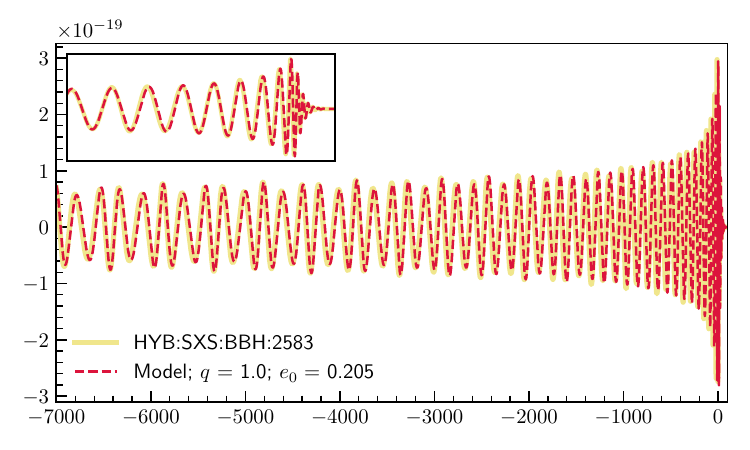}
    \includegraphics[width=0.245\textwidth]{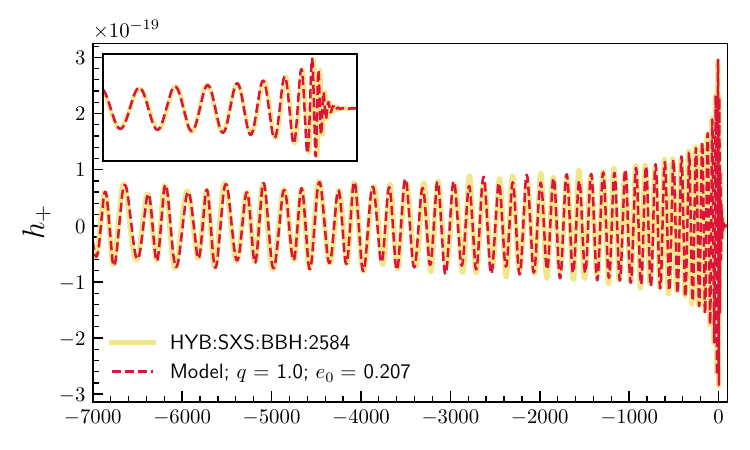}
    \includegraphics[width=0.245\textwidth]{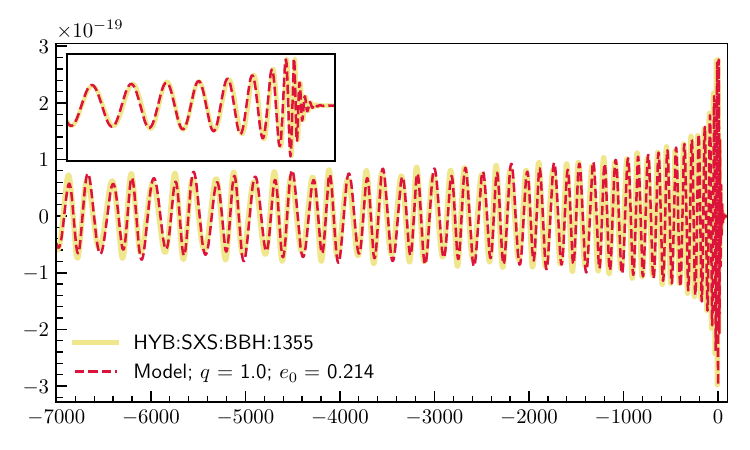}
    \includegraphics[width=0.245\textwidth]{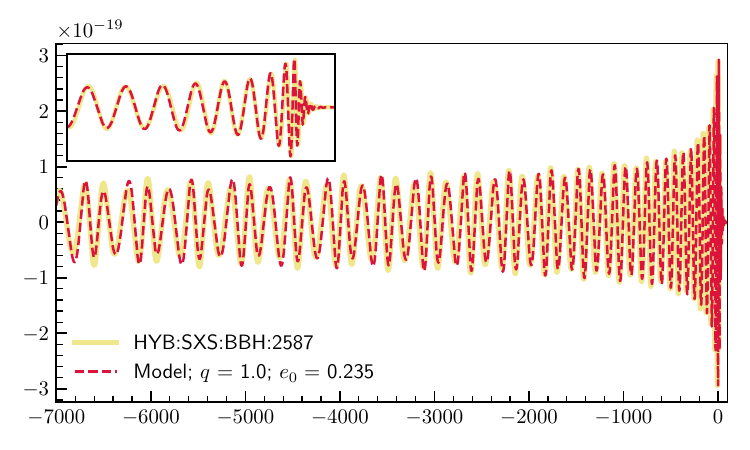}
    \includegraphics[width=0.245\textwidth]{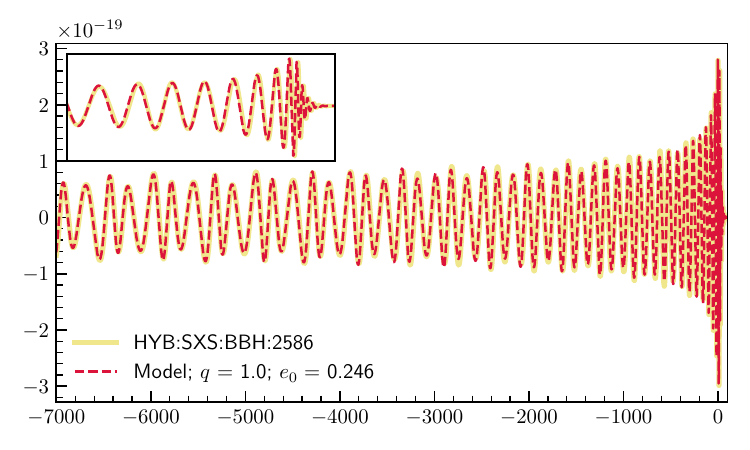}
    \includegraphics[width=0.245\textwidth]{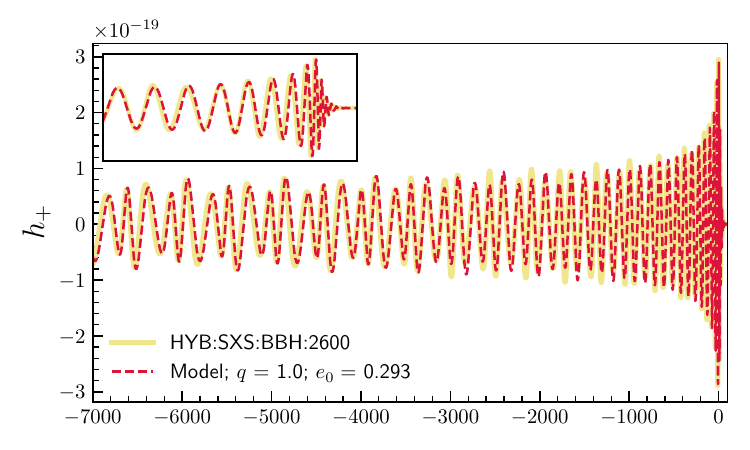}
    \includegraphics[width=0.245\textwidth]{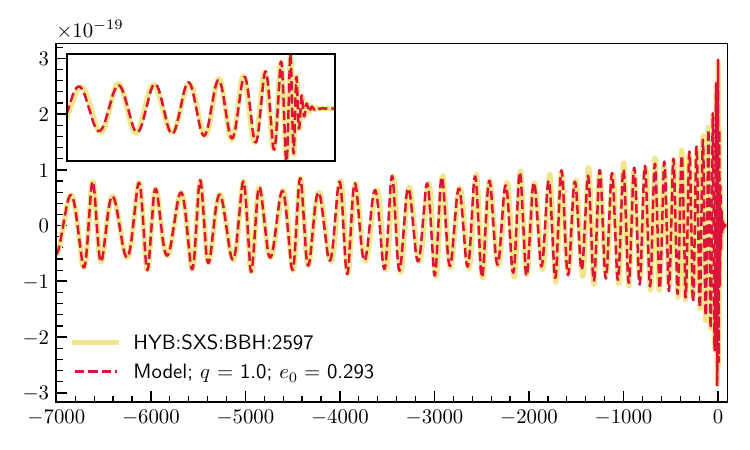}
    \includegraphics[width=0.245\textwidth]{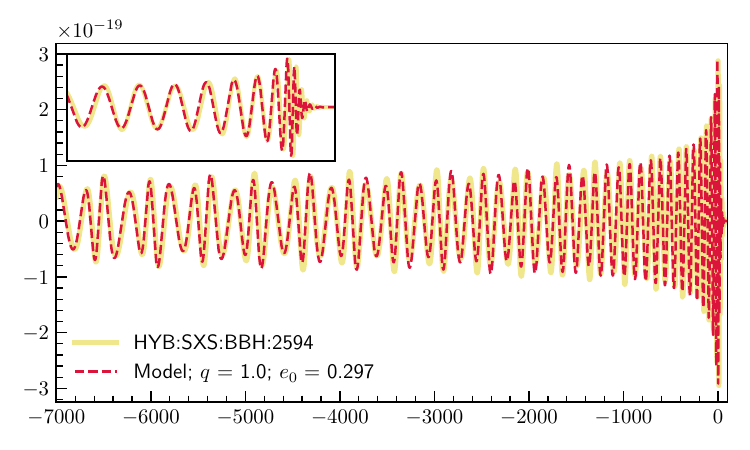}
    \includegraphics[width=0.245\textwidth]{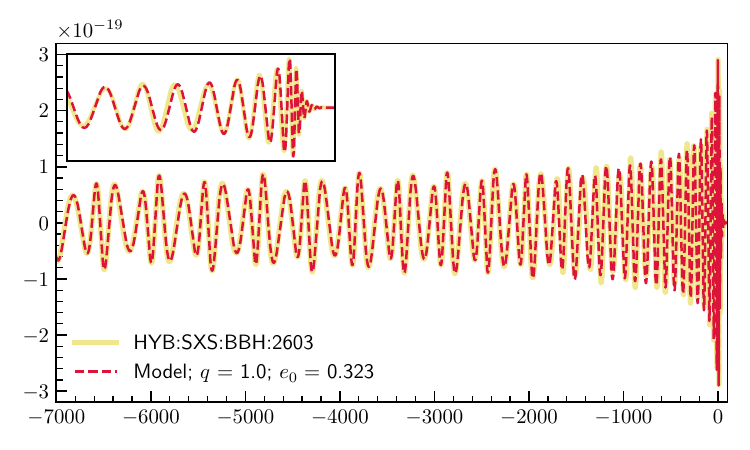}
    \includegraphics[width=0.245\textwidth]{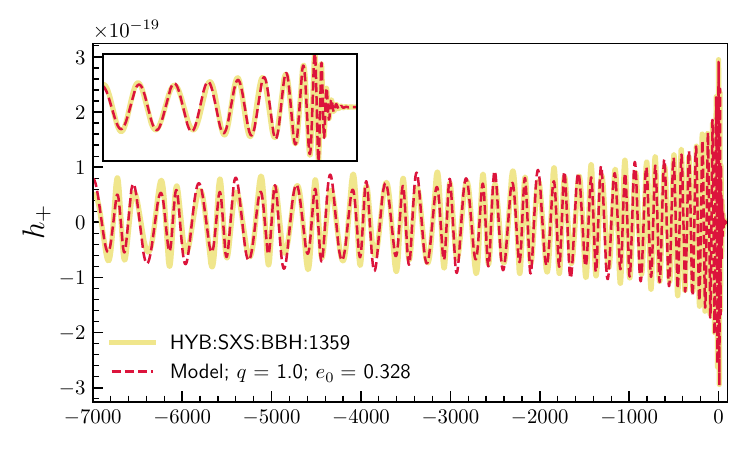}
    \includegraphics[width=0.245\textwidth]{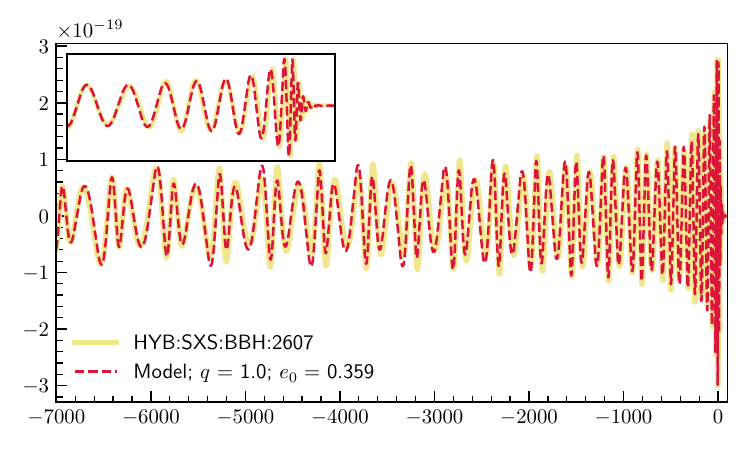}
    \includegraphics[width=0.245\textwidth]{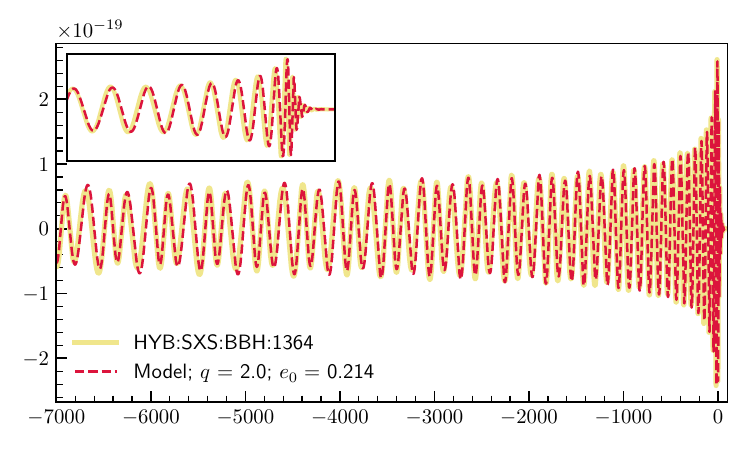}
    \includegraphics[width=0.245\textwidth]{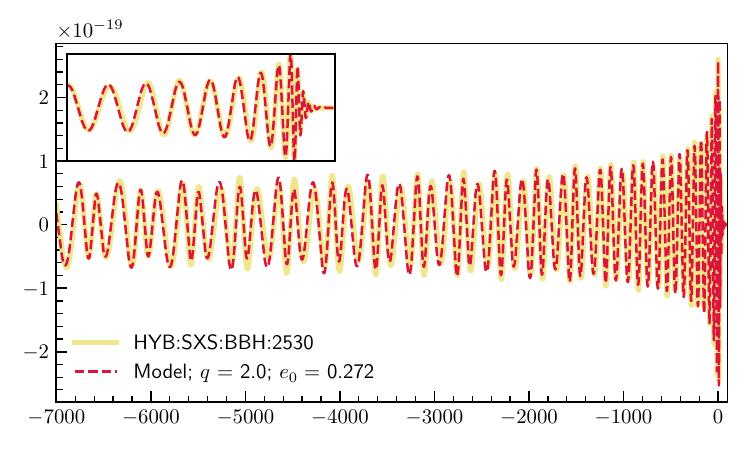}
    \includegraphics[width=0.245\textwidth]{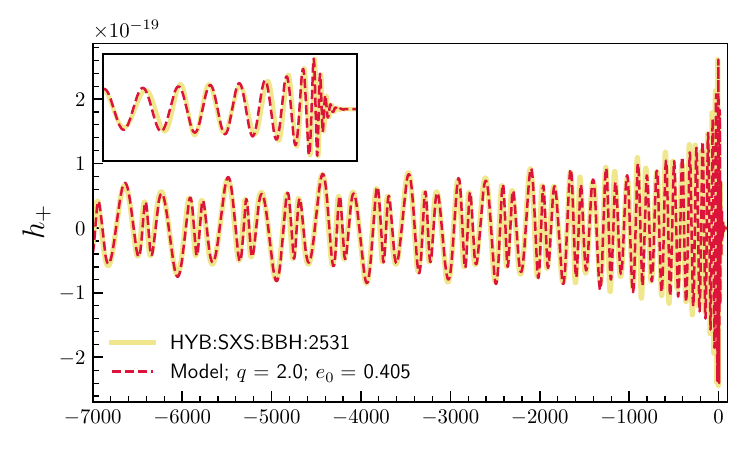}
    \includegraphics[width=0.245\textwidth]{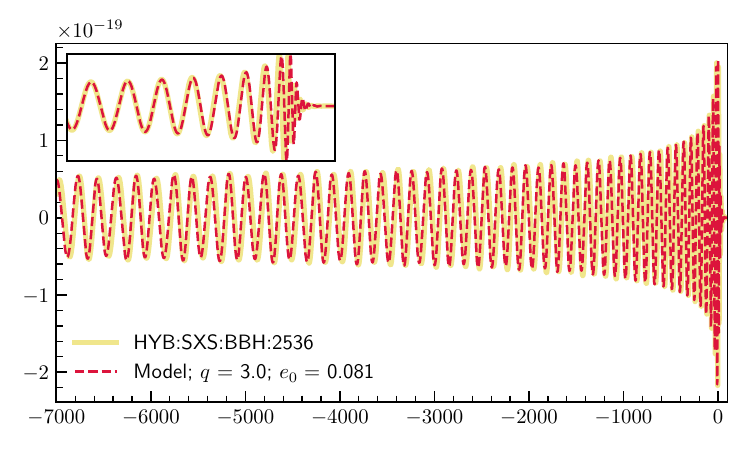}
    \includegraphics[width=0.245\textwidth]{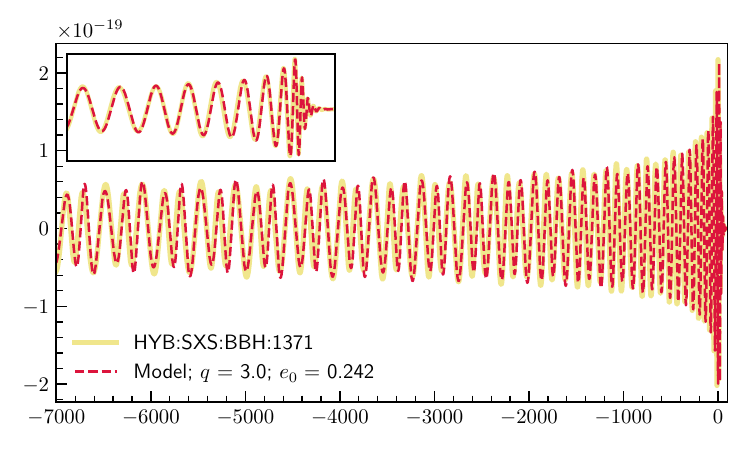}
    \includegraphics[width=0.245\textwidth]{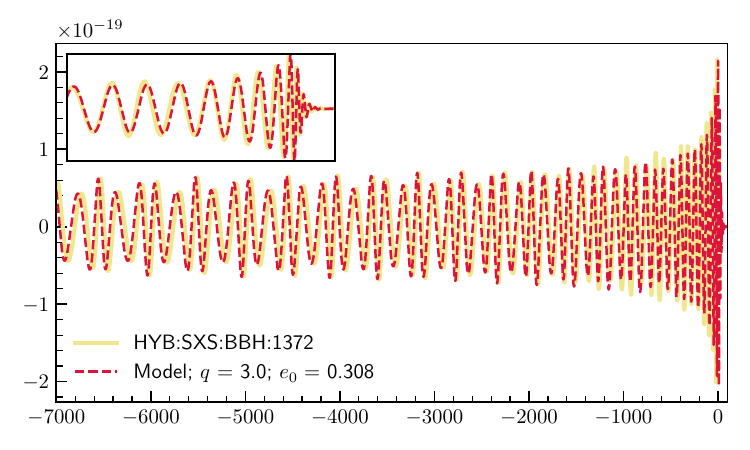}
    \includegraphics[width=0.245\textwidth]{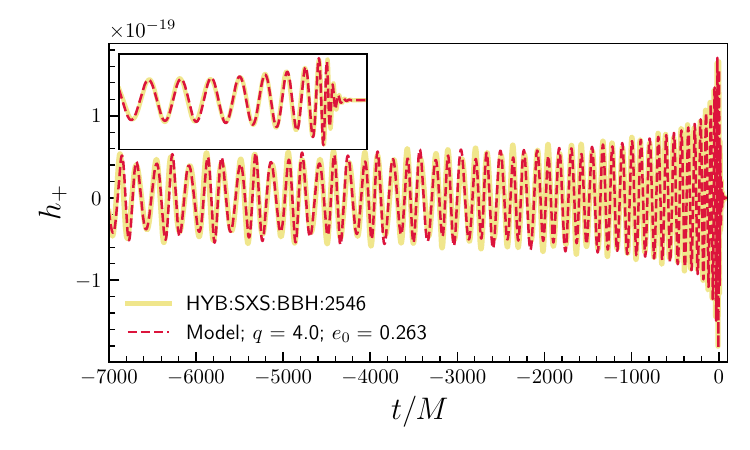}
    \includegraphics[width=0.245\textwidth]{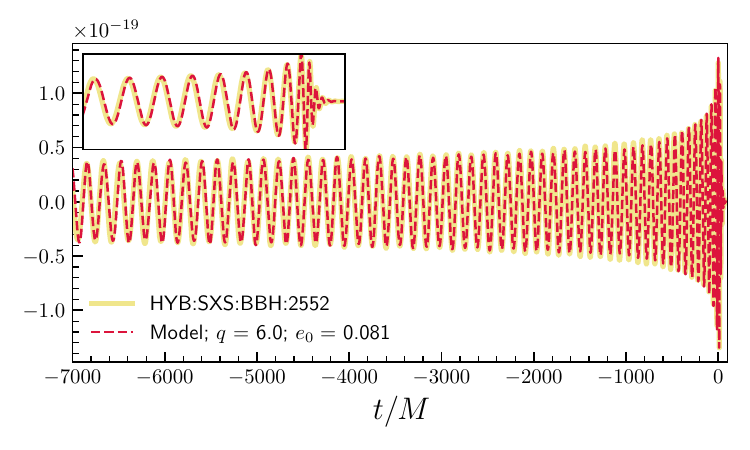}
    \includegraphics[width=0.245\textwidth]{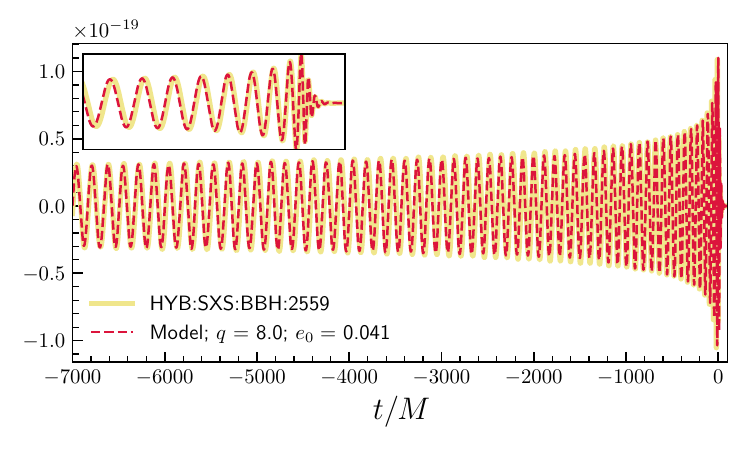}
    \includegraphics[width=0.245\textwidth]{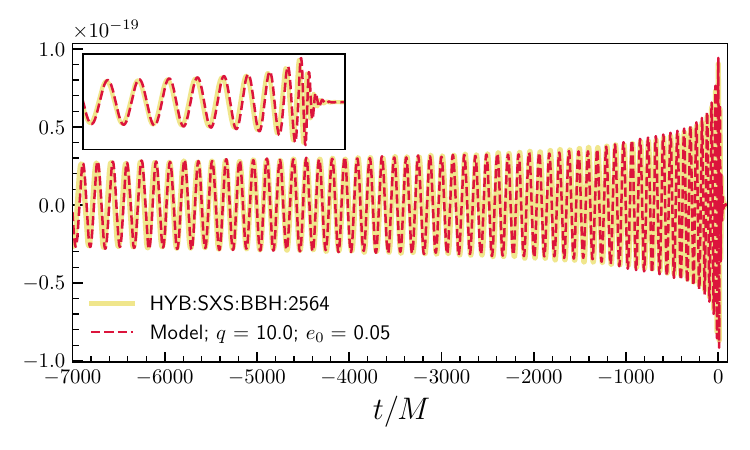}
    \caption{Comparison of strains $h_+(=\mathrm{Re}[\mathpzc{h}_{\rm 22}])$  between \texttt{EccHybIMR22} (constructed in Sec.~\ref{sec:22-mode-model}) and the \textit{training set} hybrids (listed in Table~\ref{tab:nrsims}) for a total mass of $40M_\odot$. The inset shows an enlarged view of the late inspiral and merger-ringdown cycles.}
    \label{strain_comp_training}
\end{figure*}

\begin{figure*}[htbp!]
\centering
    \includegraphics[width=0.24\textwidth]{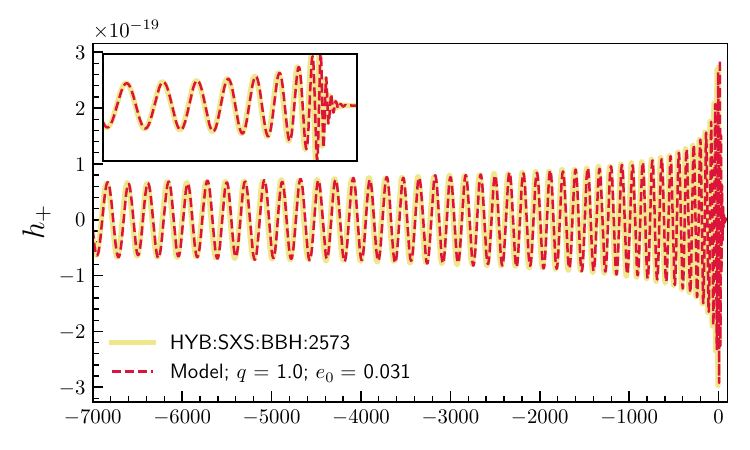}
    \includegraphics[width=0.24\textwidth]{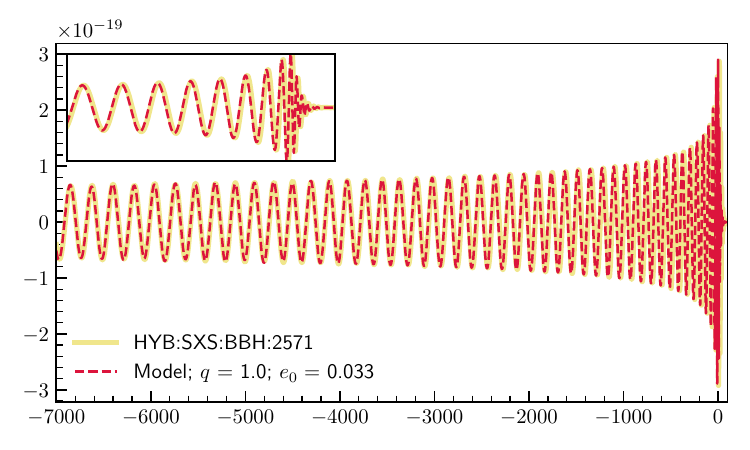}
    \includegraphics[width=0.24\textwidth]{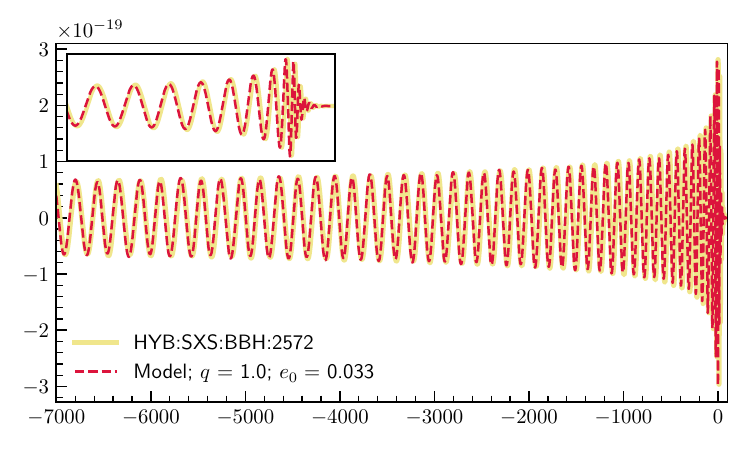}
    \includegraphics[width=0.24\textwidth]{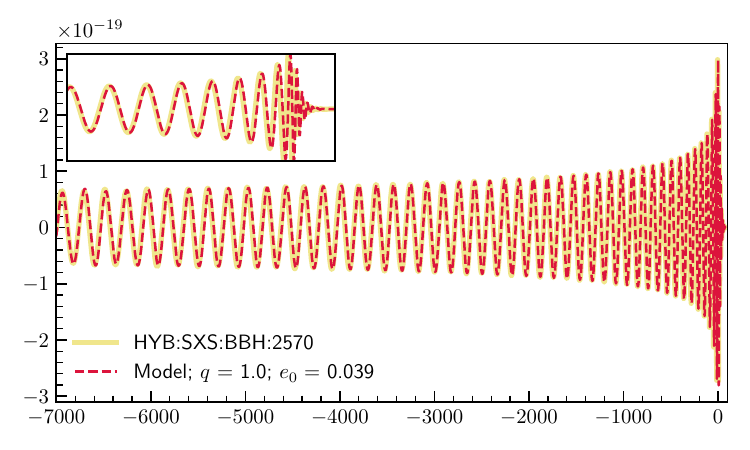}
    \includegraphics[width=0.24\textwidth]{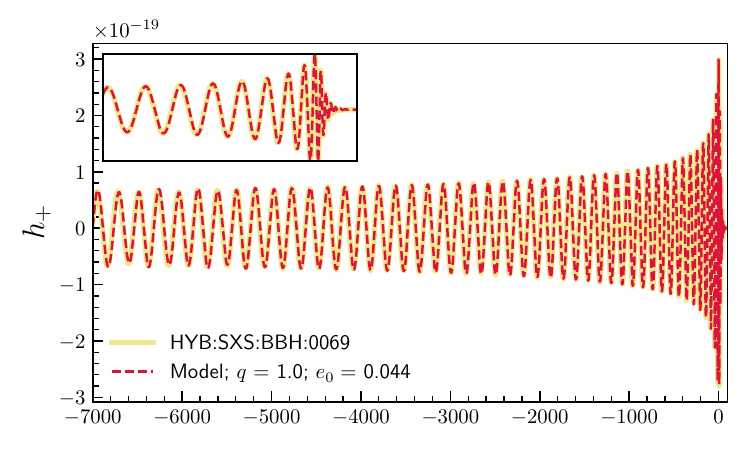}
    \includegraphics[width=0.24\textwidth]{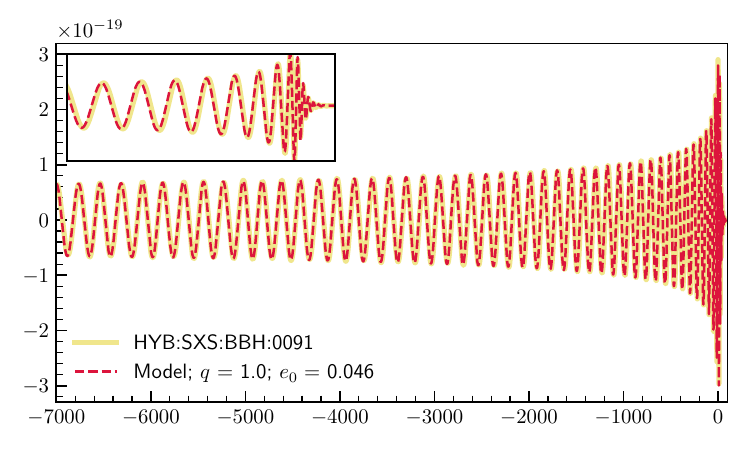}
    \includegraphics[width=0.24\textwidth]{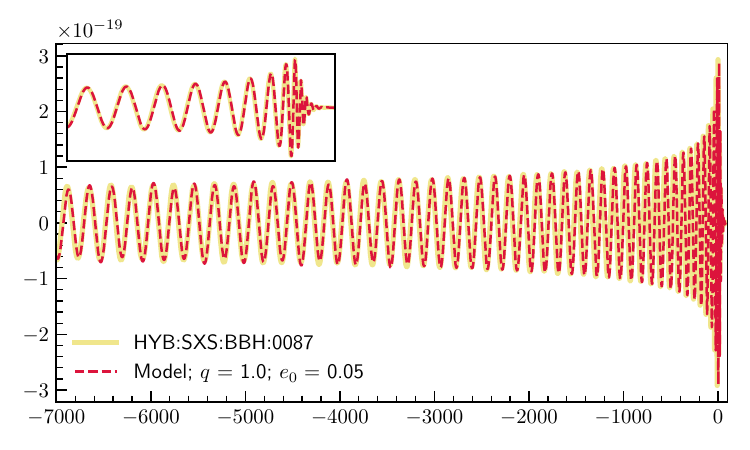}
    \includegraphics[width=0.24\textwidth]{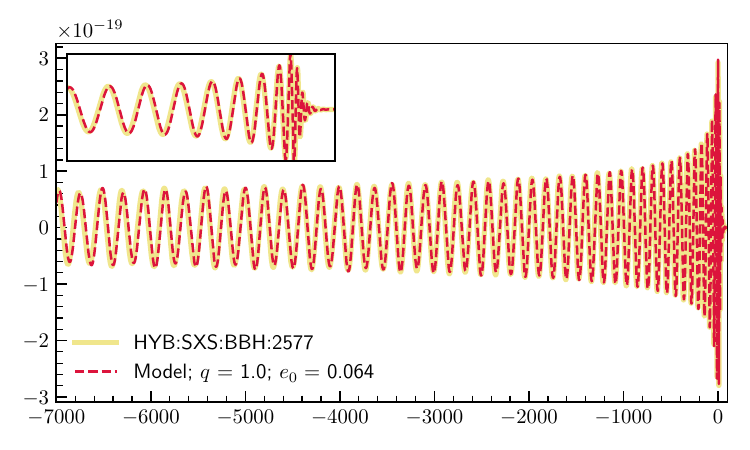}
    \includegraphics[width=0.24\textwidth]{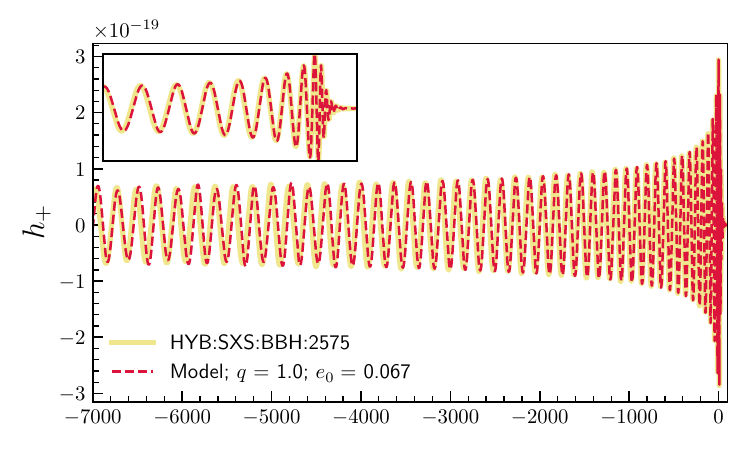}
    \includegraphics[width=0.24\textwidth]{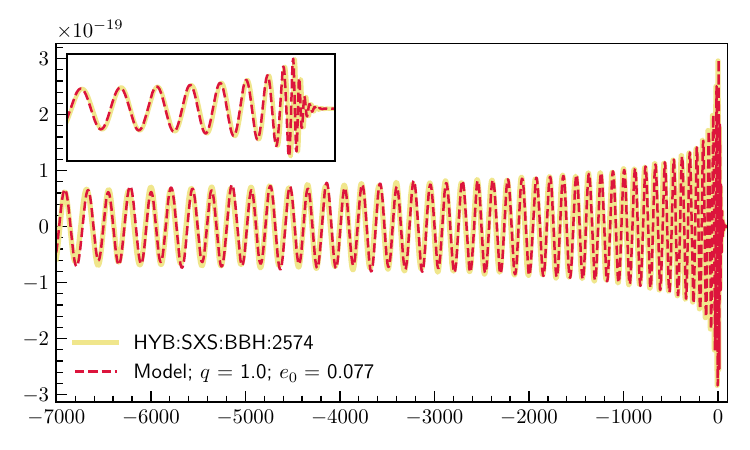}
    \includegraphics[width=0.24\textwidth]{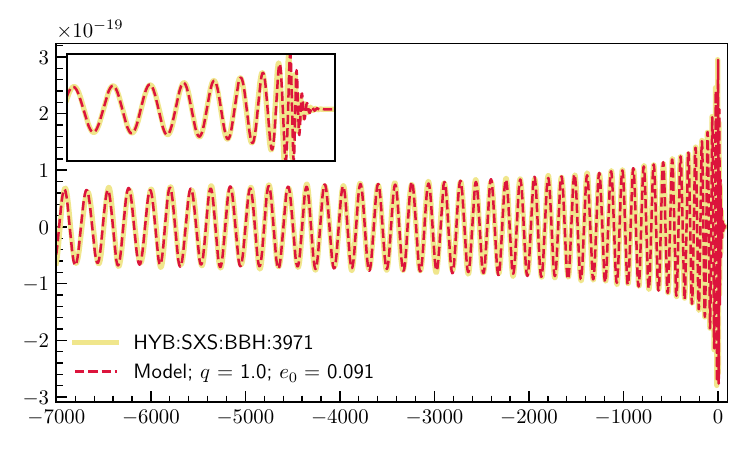}
    \includegraphics[width=0.24\textwidth]{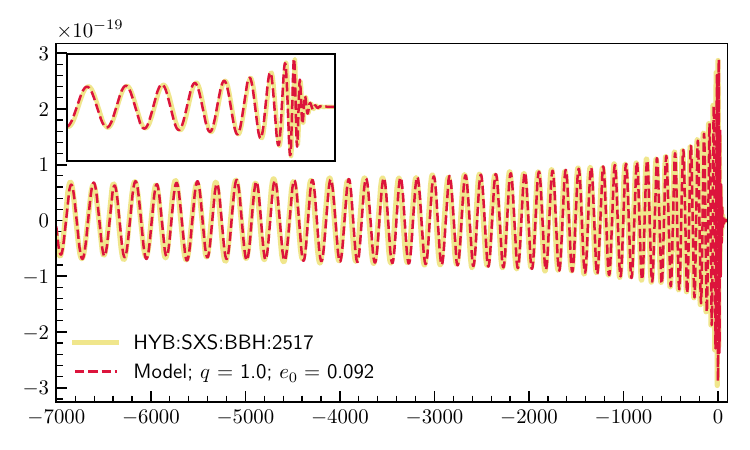}
    \includegraphics[width=0.24\textwidth]{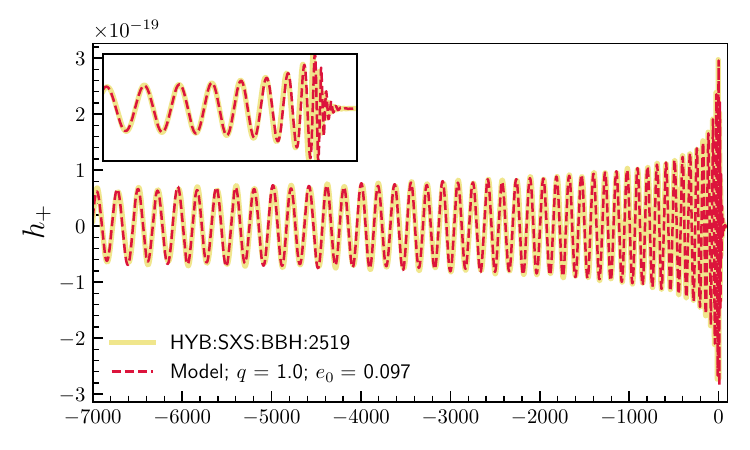}
    \includegraphics[width=0.24\textwidth]{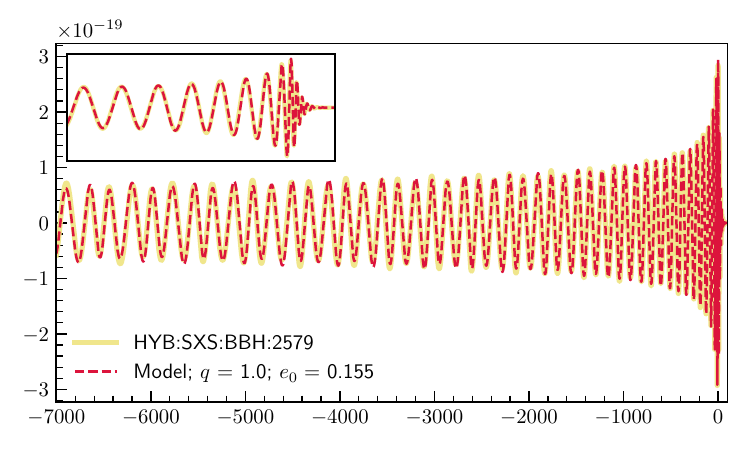}
    \includegraphics[width=0.24\textwidth]{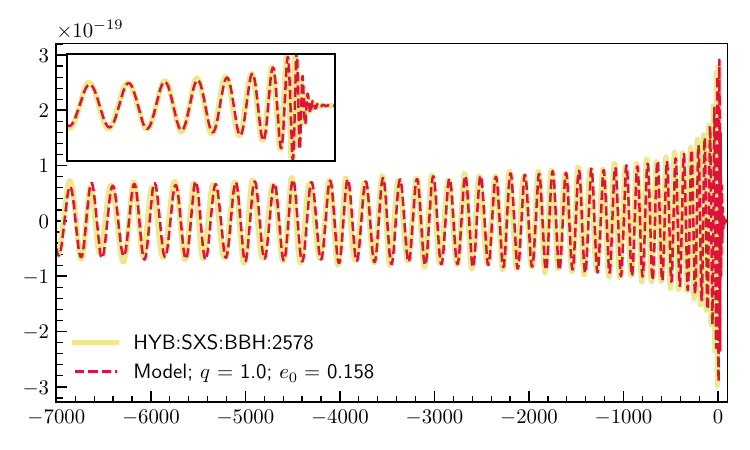}
    \includegraphics[width=0.24\textwidth]{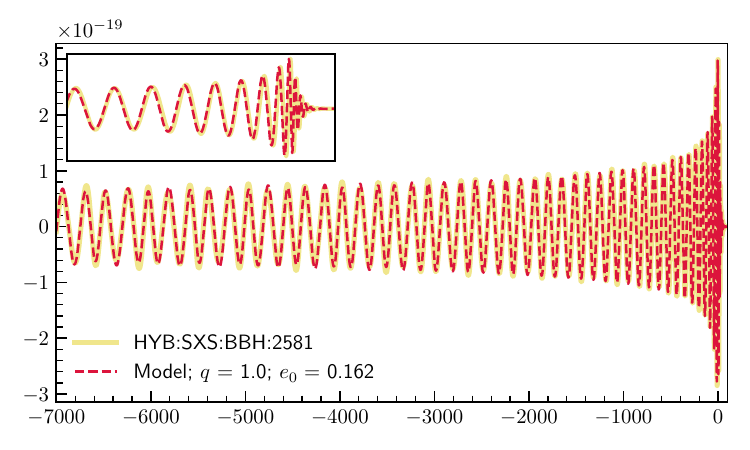}
    \includegraphics[width=0.24\textwidth]{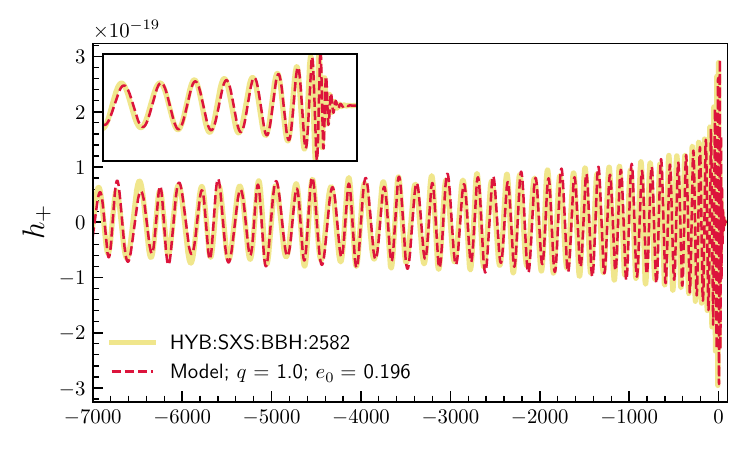}
    \includegraphics[width=0.24\textwidth]{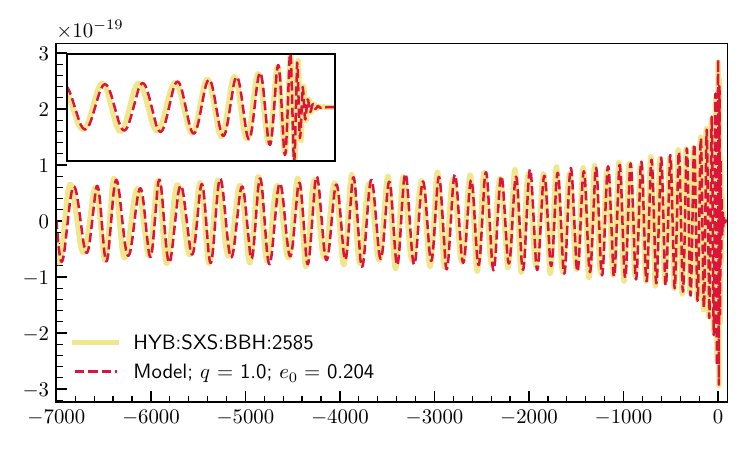}
    \includegraphics[width=0.24\textwidth]{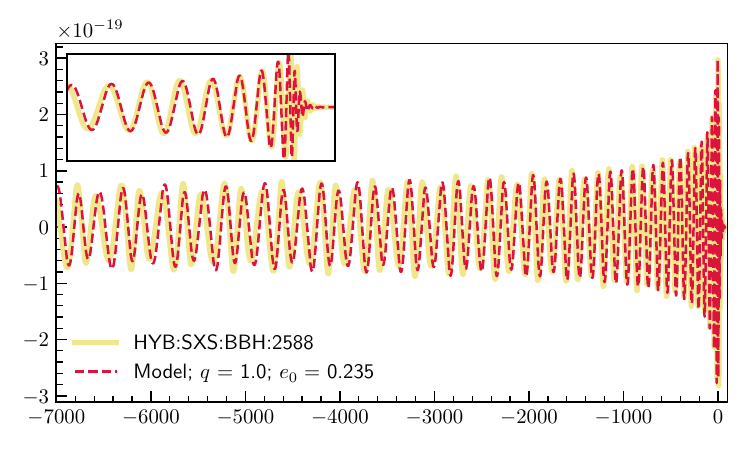}
    \includegraphics[width=0.24\textwidth]{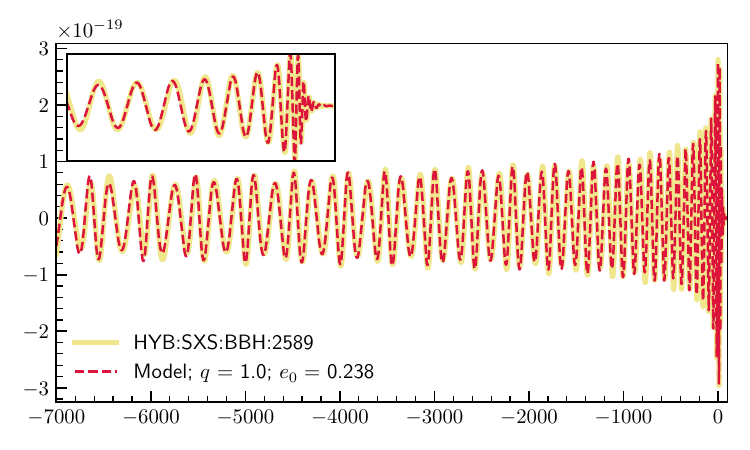}
    \includegraphics[width=0.24\textwidth]{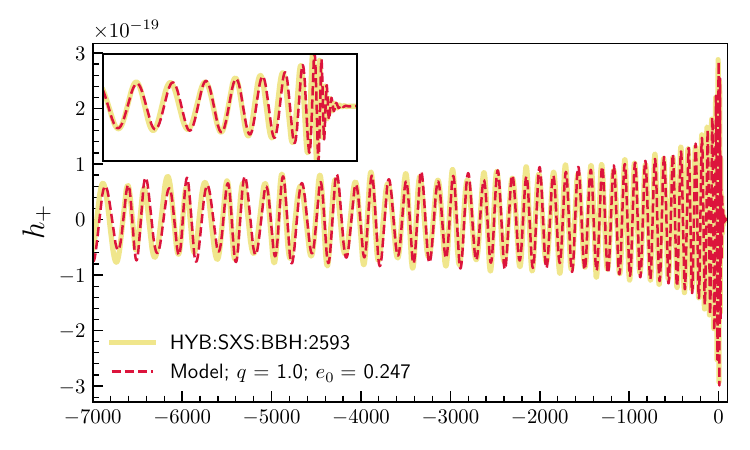}
    \includegraphics[width=0.24\textwidth]{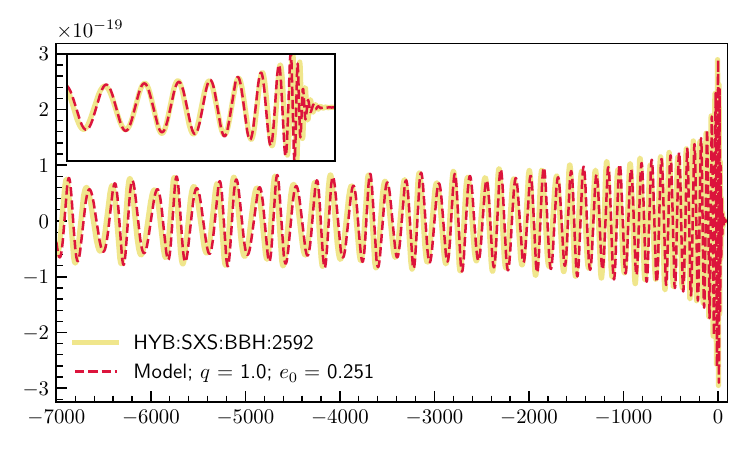}
    \includegraphics[width=0.24\textwidth]{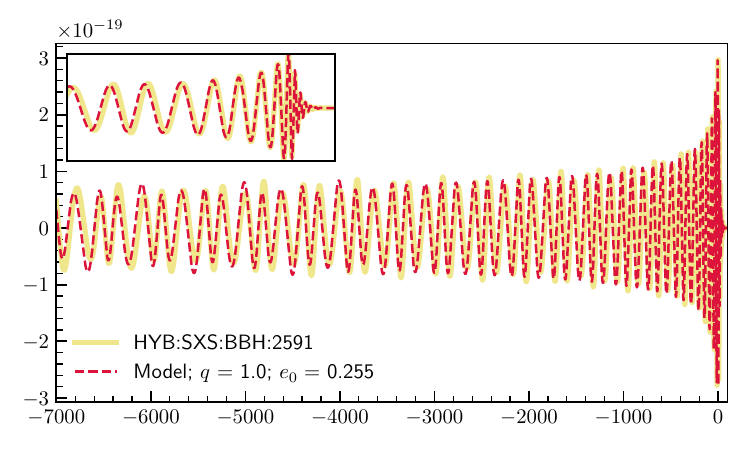}
    \includegraphics[width=0.24\textwidth]{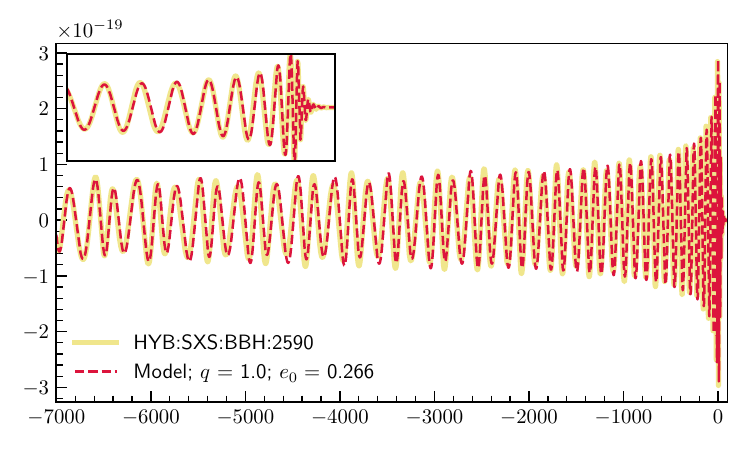}
    \includegraphics[width=0.24\textwidth]{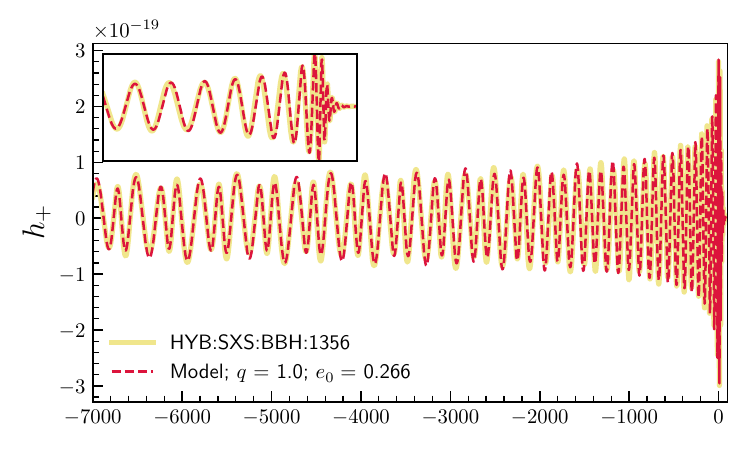}
    \includegraphics[width=0.24\textwidth]{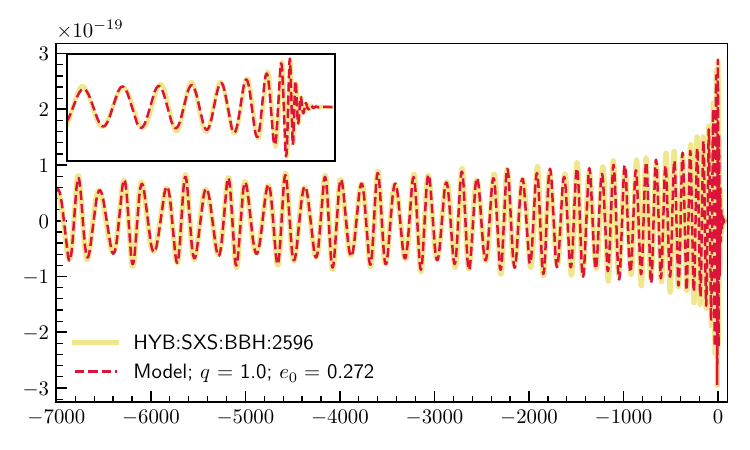}
    \includegraphics[width=0.24\textwidth]{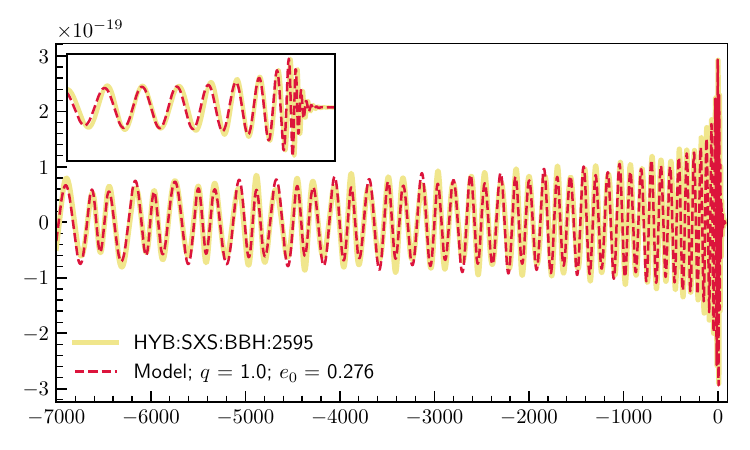}
    \includegraphics[width=0.24\textwidth]{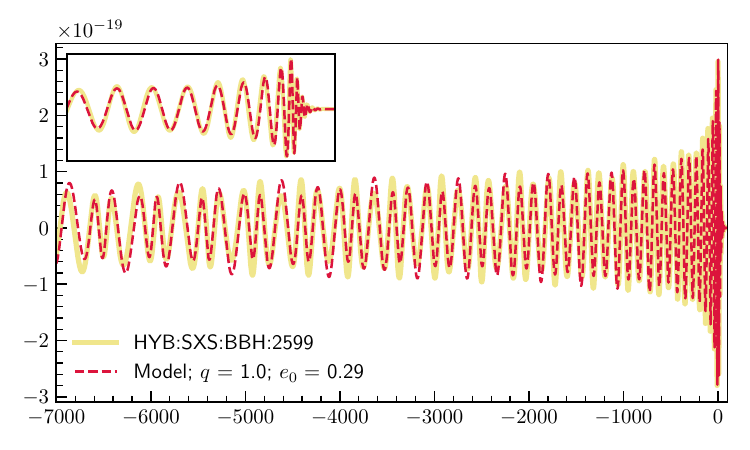}
    \includegraphics[width=0.24\textwidth]{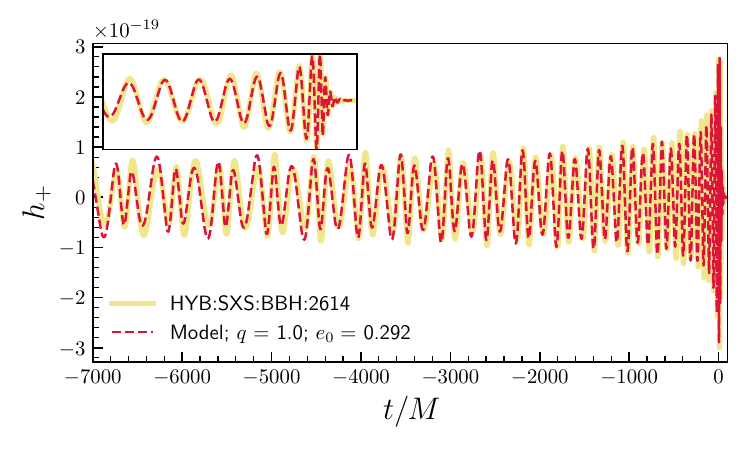}
    \includegraphics[width=0.24\textwidth]{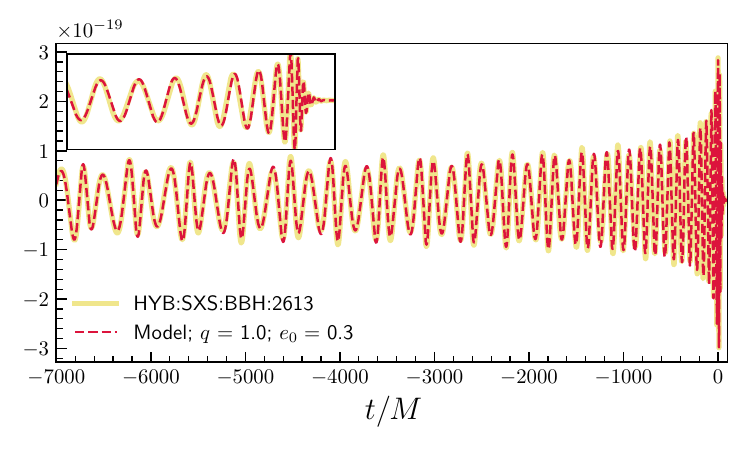}
    \includegraphics[width=0.24\textwidth]{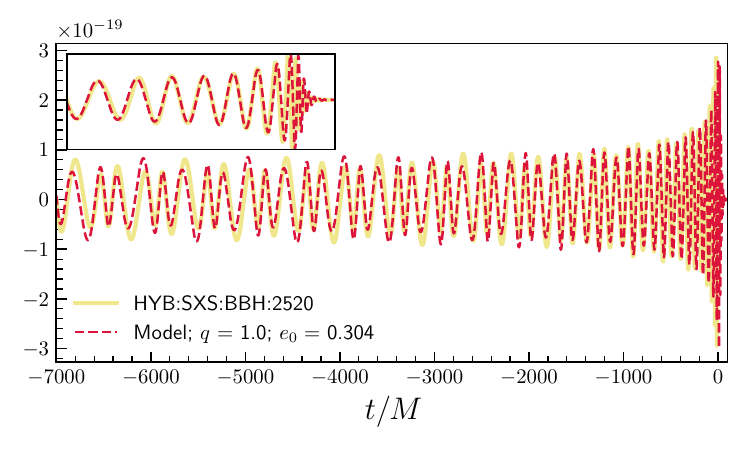}
    \includegraphics[width=0.24\textwidth]{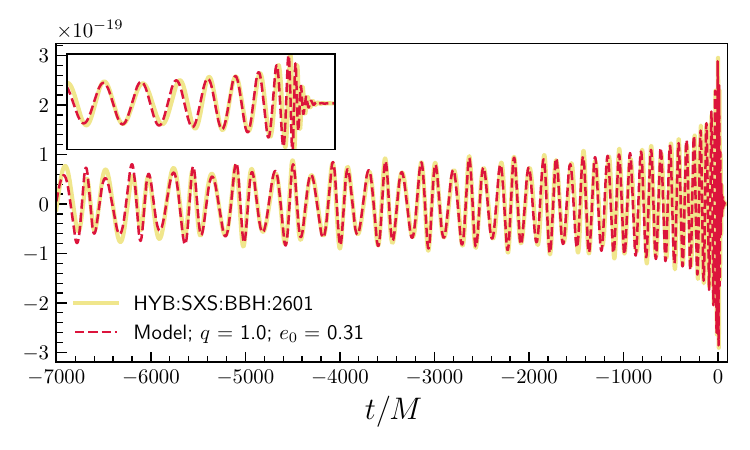}
    \caption{Same as Fig.~\ref{strain_comp_training}, except the comparison is done against hybrids taken from the \textit{testing set} (listed in Table~\ref{tab:nrsims}).}
    \label{strain_comp_testing}
\end{figure*}

\addtocounter{figure}{-1}
\begin{figure*}[htbp!]
\centering

    \includegraphics[width=0.24\textwidth]{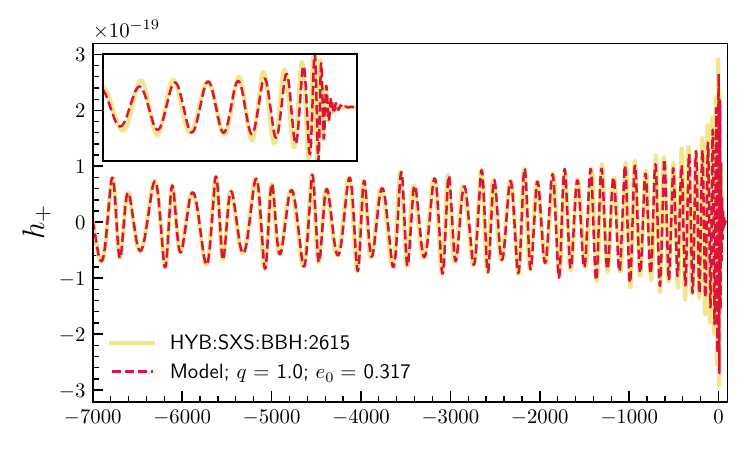}
    \includegraphics[width=0.24\textwidth]{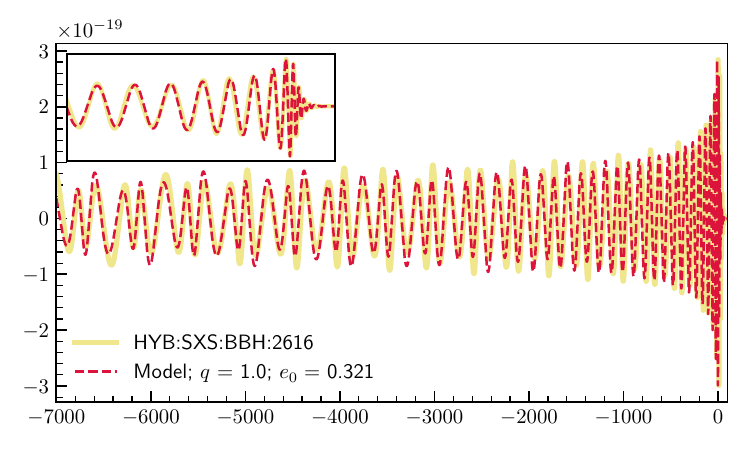}
    \includegraphics[width=0.24\textwidth]{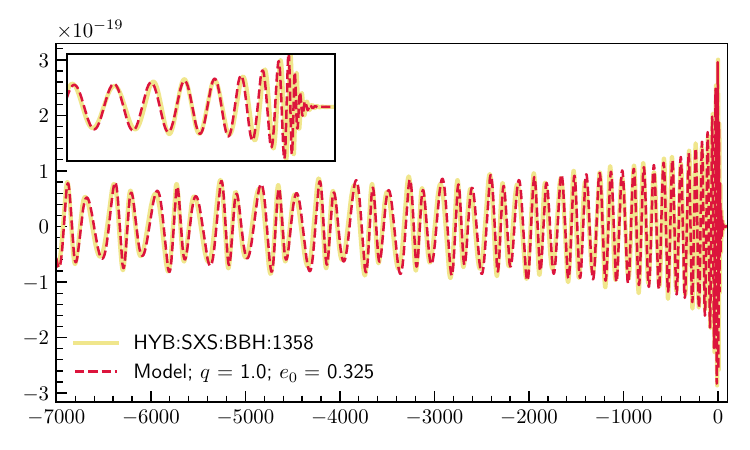}
    \includegraphics[width=0.24\textwidth]{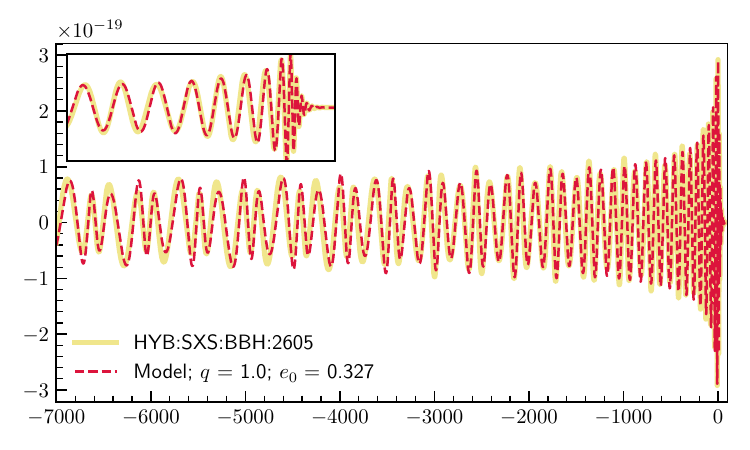}
    \includegraphics[width=0.24\textwidth]{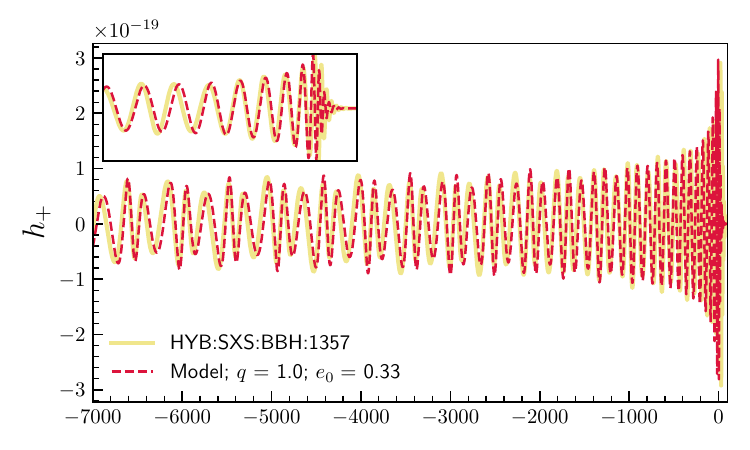}
    \includegraphics[width=0.24\textwidth]{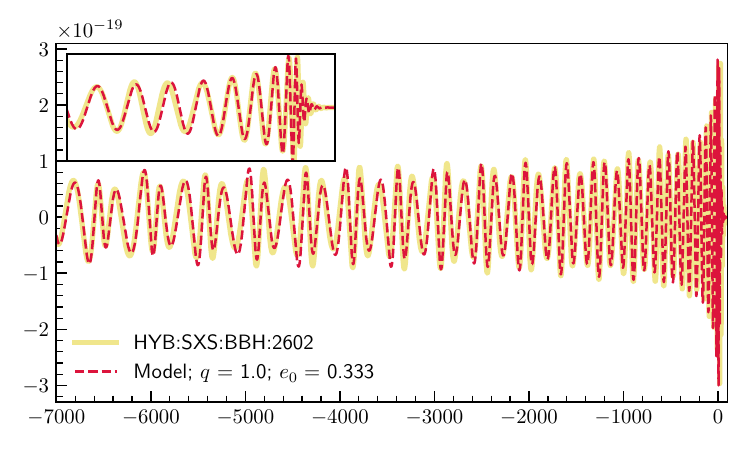}
    \includegraphics[width=0.24\textwidth]{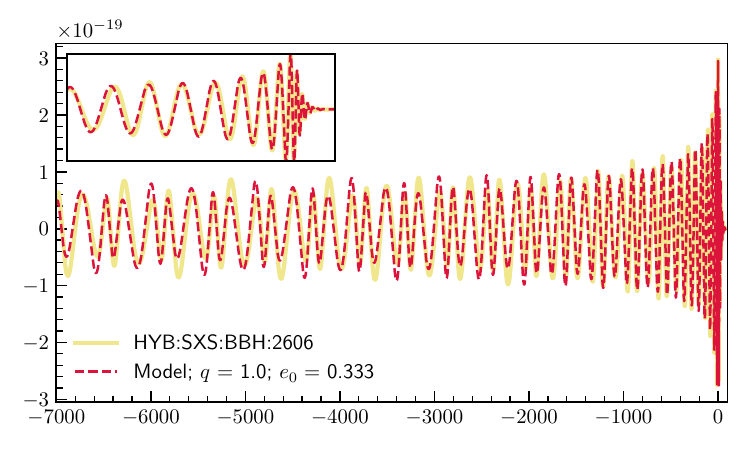}
    \includegraphics[width=0.24\textwidth]{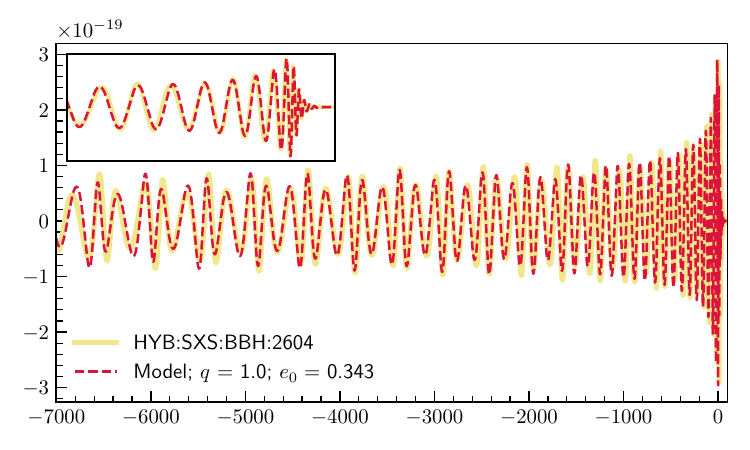}
    \includegraphics[width=0.24\textwidth]{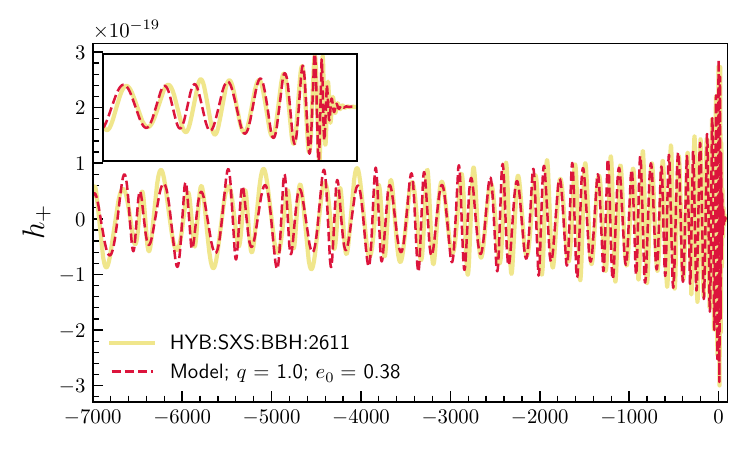}
    \includegraphics[width=0.24\textwidth]{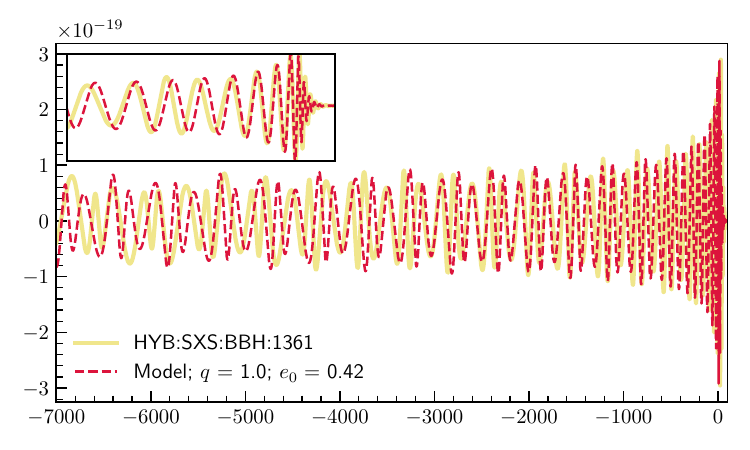}
    \includegraphics[width=0.24\textwidth]{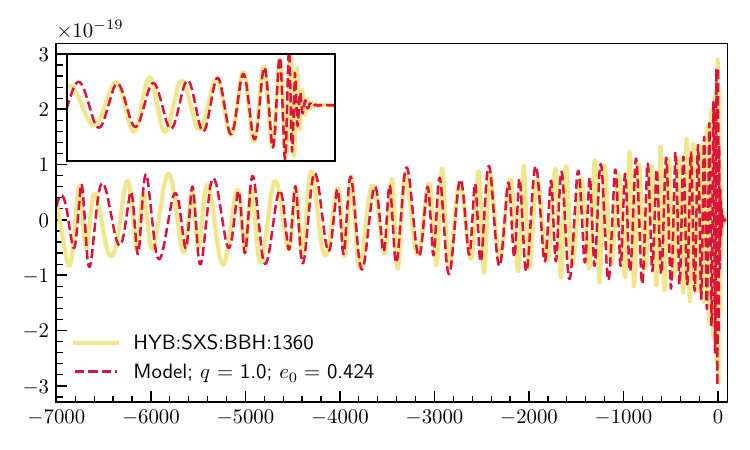}
    \includegraphics[width=0.24\textwidth]{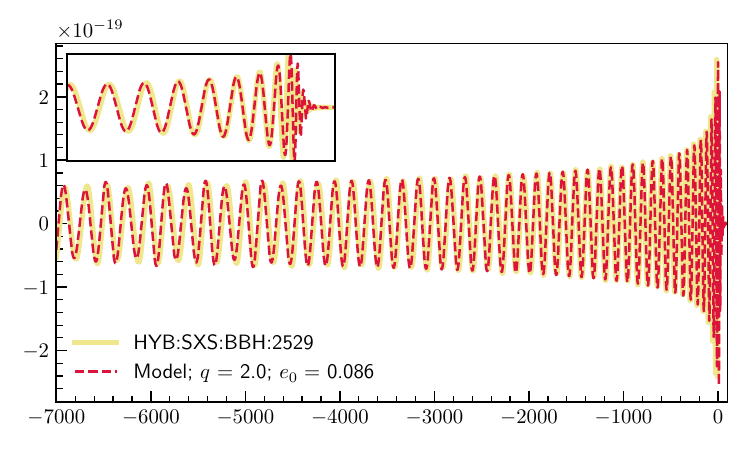}
    \includegraphics[width=0.24\textwidth]{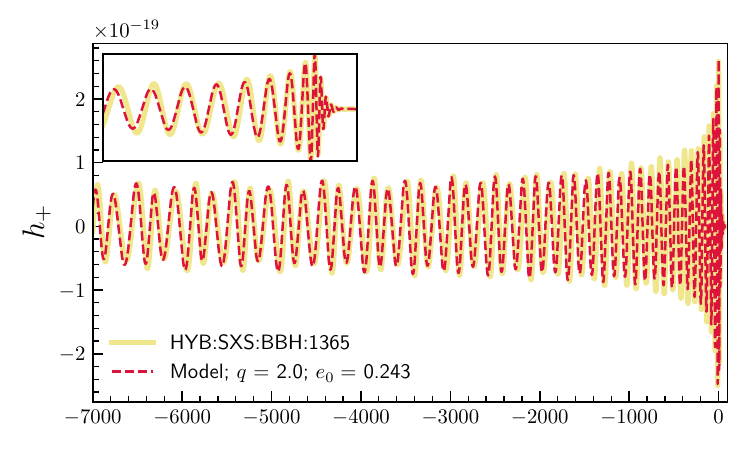}
    \includegraphics[width=0.24\textwidth]{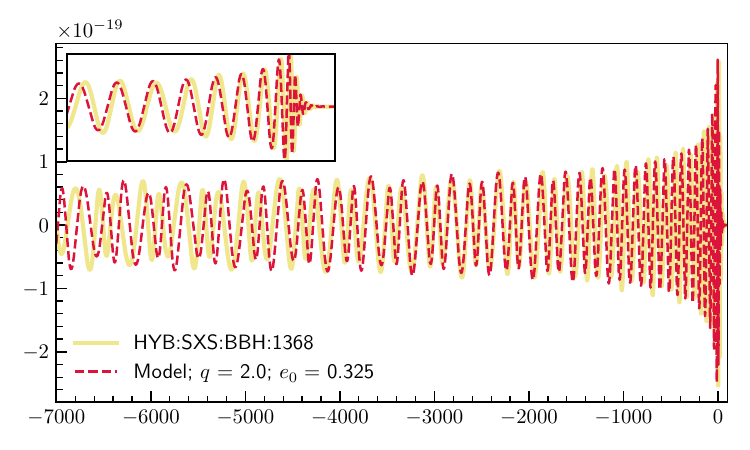}
    \includegraphics[width=0.24\textwidth]{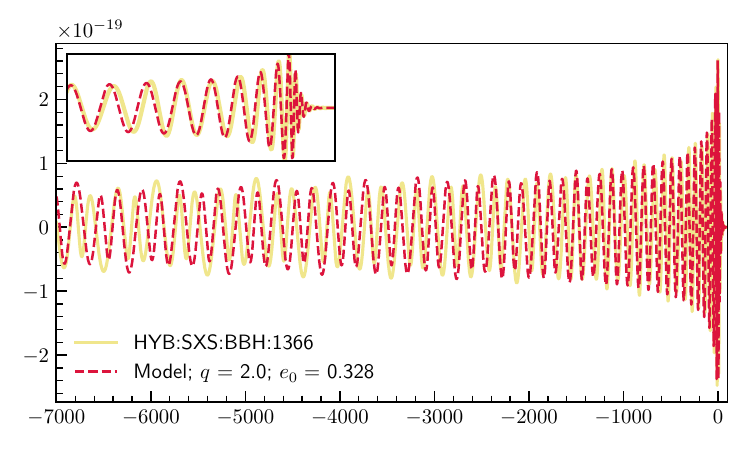}
    \includegraphics[width=0.24\textwidth]{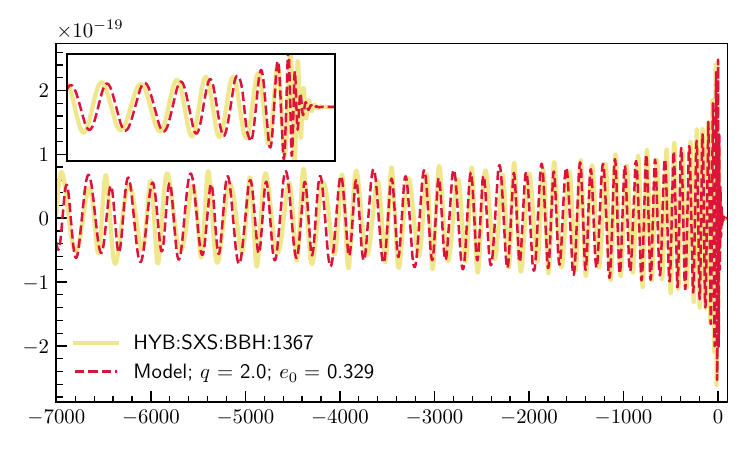}
    \includegraphics[width=0.24\textwidth]{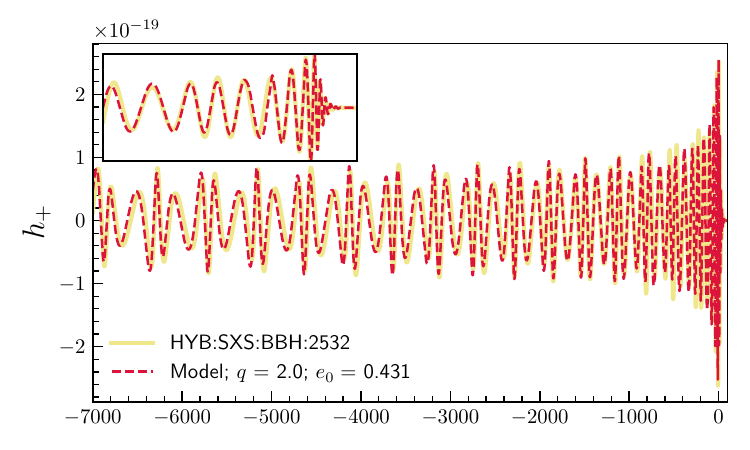}
    \includegraphics[width=0.24\textwidth]{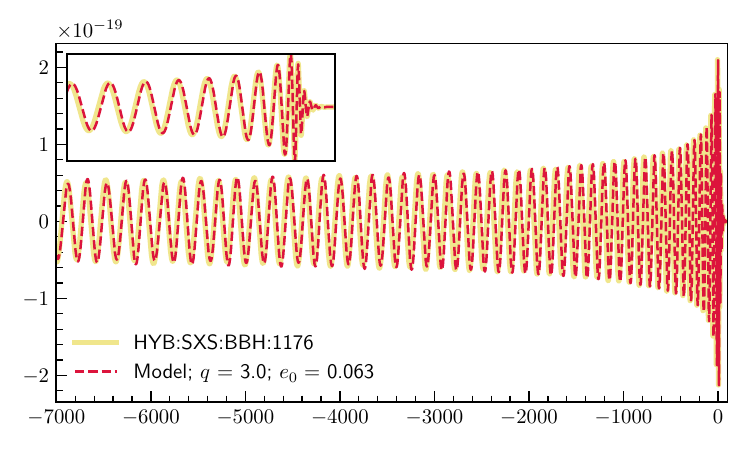}
    \includegraphics[width=0.24\textwidth]{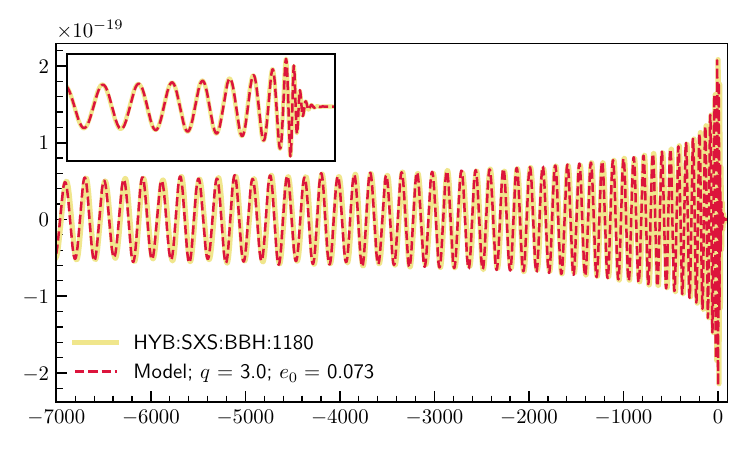}
    \includegraphics[width=0.24\textwidth]{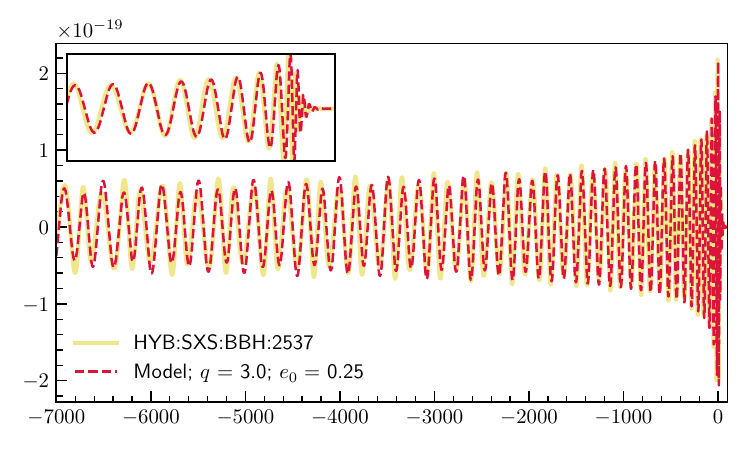}
    \includegraphics[width=0.24\textwidth]{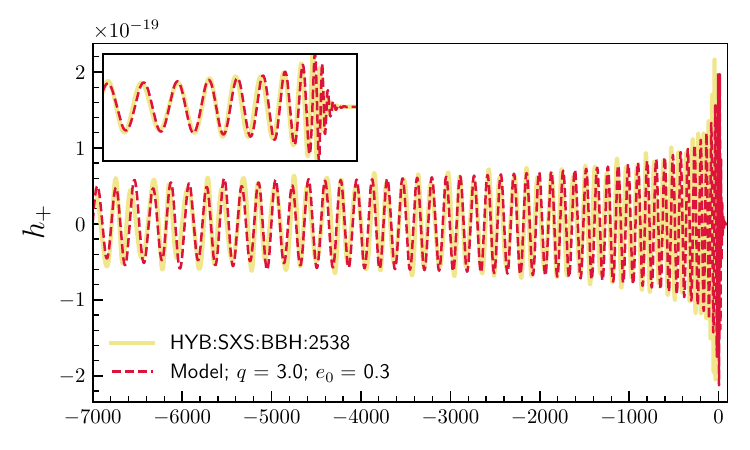}
    \includegraphics[width=0.24\textwidth]{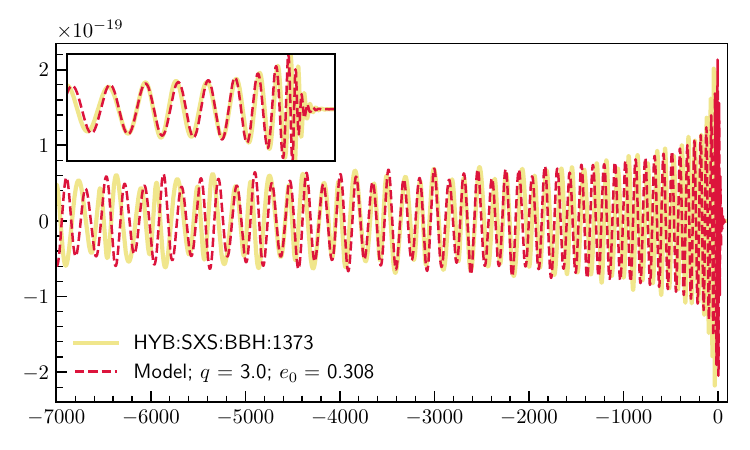}
    \includegraphics[width=0.24\textwidth]{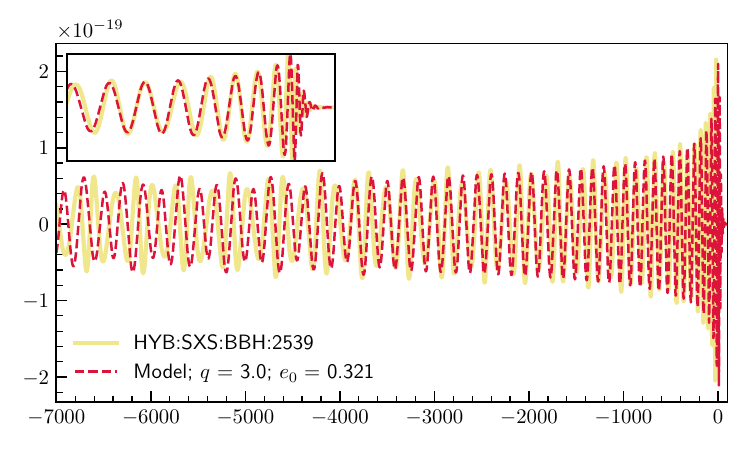}
    \includegraphics[width=0.24\textwidth]{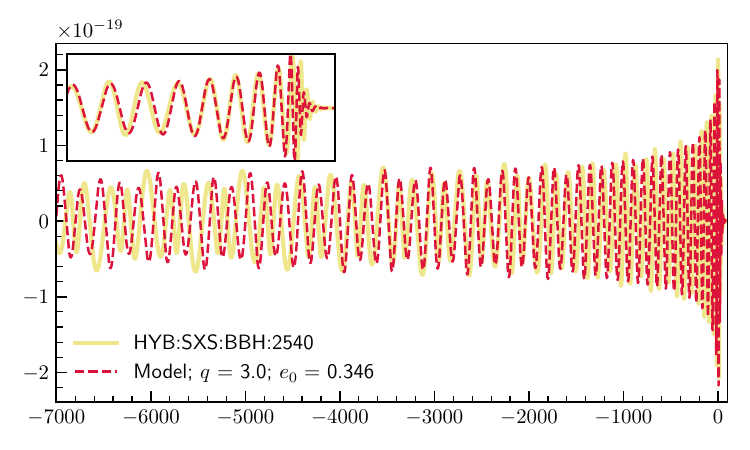}
    \includegraphics[width=0.24\textwidth]{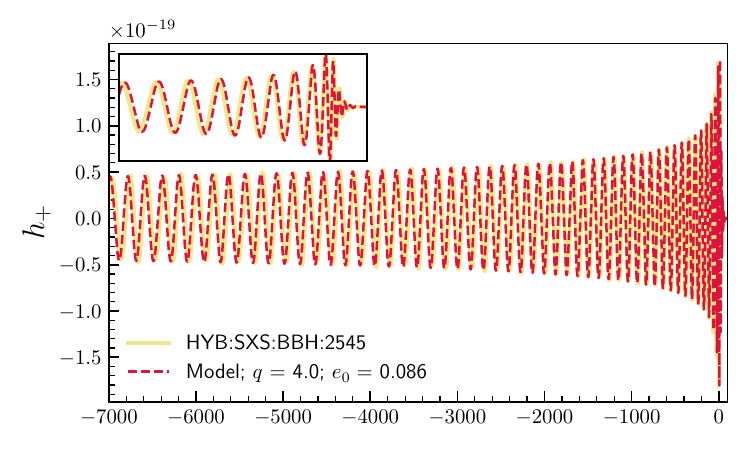}
    \includegraphics[width=0.24\textwidth]{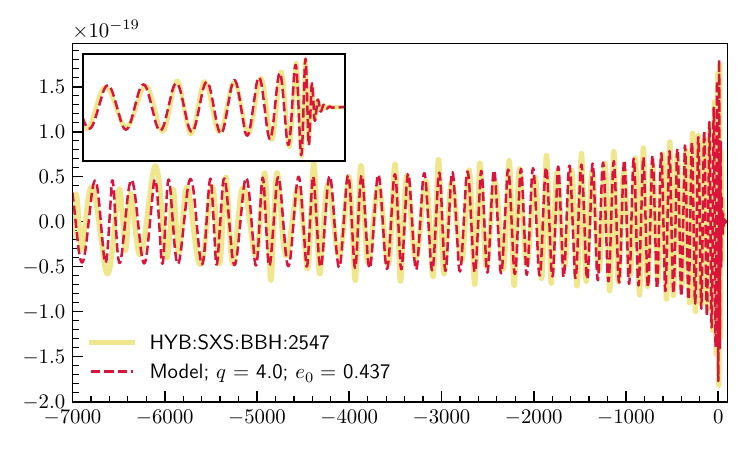}
    \includegraphics[width=0.24\textwidth]{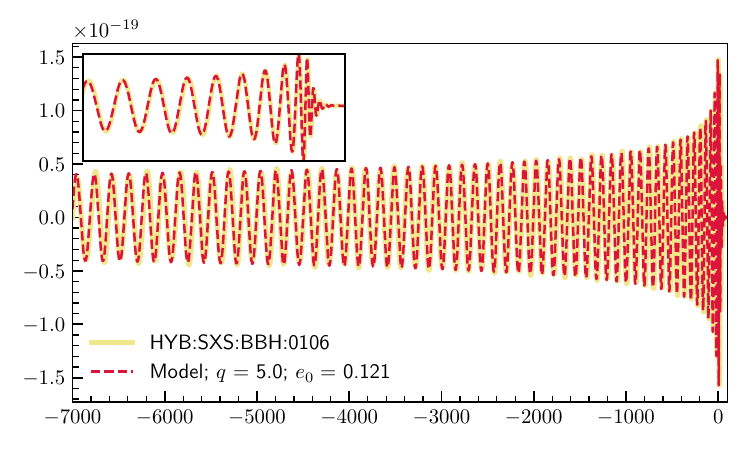}
    \includegraphics[width=0.24\textwidth]{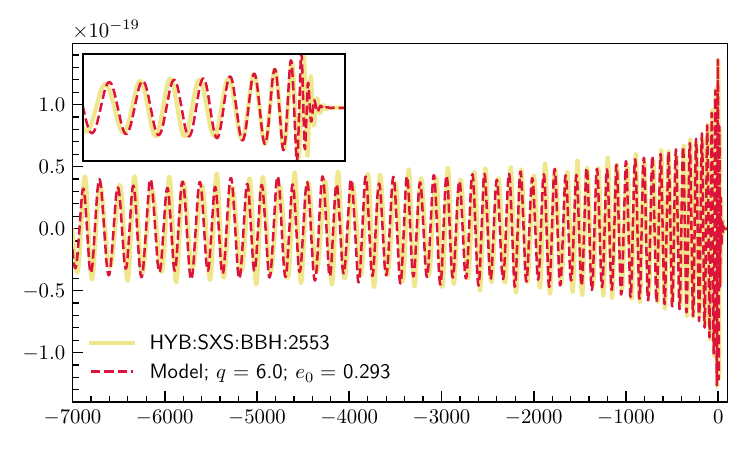}
    \includegraphics[width=0.24\textwidth]{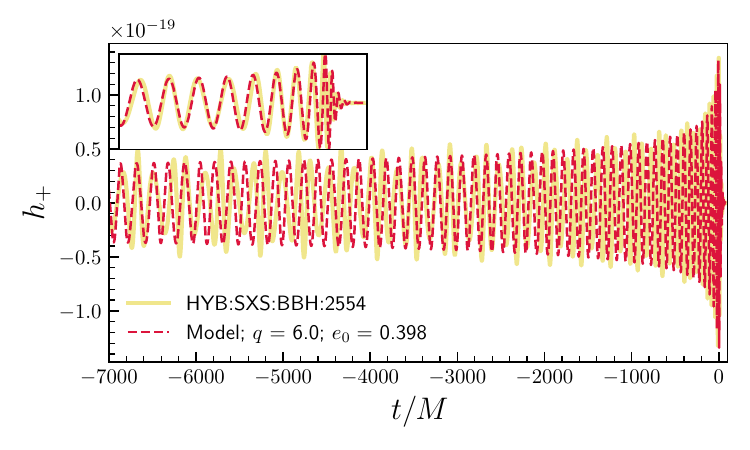}
    \includegraphics[width=0.24\textwidth]{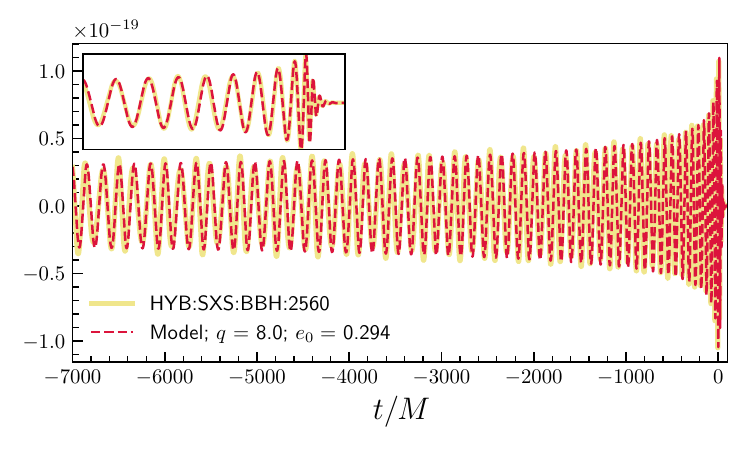}
    \includegraphics[width=0.24\textwidth]{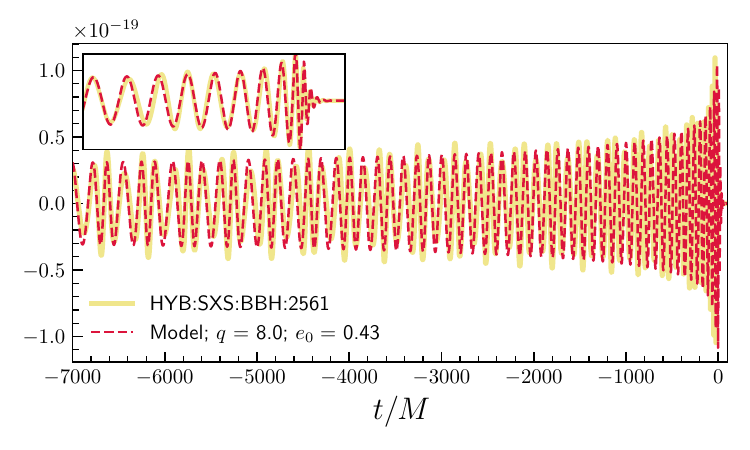}
    \includegraphics[width=0.24\textwidth]{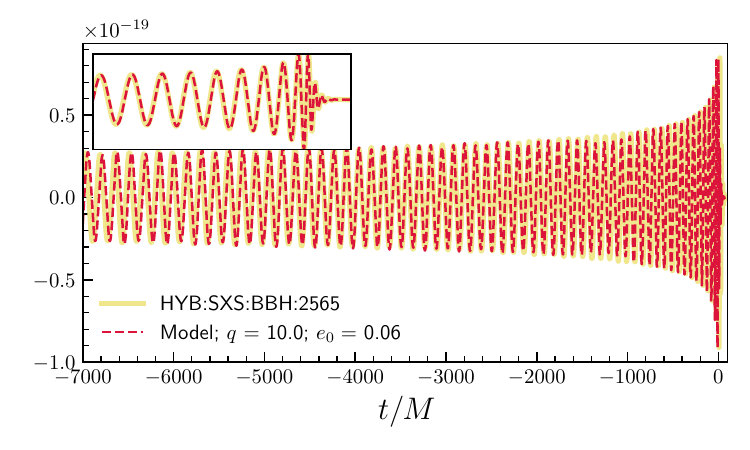}

    \caption{\textit{Continued.}}
    \label{strain_comp_testing}
\end{figure*}

\begin{table} 
\centering
\caption{Model fitting coefficients for $t_{\rm shift}$.}
\label{tab:tshift-table}
\setlength{\tabcolsep}{10pt}   
\begin{tabular}{|c|c| r r r r|}
\toprule
\hline
$k$ & $\alpha\beta$ & $A_{k\alpha\beta}$ & $\gamma_k$ & $\delta_k$ & $a_k$ \\
\midrule
\hline
 & 00 &  55.81  & &  & \\
0 & 10 & -842.9   &  6.145  & -36.20  & -1.458 \\
 & 01 & -43.04  & & & \\[4pt]
 & 00 & -251.5    &    &  &   \\
1 & 10 &  69.35   &  2.957   & -17.83  &  1.397 \\
 & 01 & 106.1    &   &   &  \\[4pt]
 & 00 & -304.7   &   &    &   \\
2 & 10 & 1170    & -3.622  &  8.608  &  0.027 \\
 & 01 &  34.88   &   &   &   \\[4pt]
 & 00 & -114.3   &   &    &   \\
3 & 10 &  592.7    & -0.438  &  4.819  &  0.403 \\
 & 01 &  69.43    &   &    &   \\
\hline
\bottomrule
\end{tabular}
\end{table}

\begin{table} 
\centering
\caption{Model fitting coefficients for  $t^\mathcal{A}_{\rm match}(q,e,l)$.}
\label{tab:tamp-table}
\setlength{\tabcolsep}{18pt}   
\begin{tabular}{|c|c| r r|}
\toprule
\hline
$k$ & $\alpha\beta$ & $A_{k\alpha\beta}\ (\times10^3)$ & $a_k$ \\
\hline
\midrule
\multirow{3}{*}{0}  
& 00 &  -1.990 & \multirow{3}{*}{---} \\
& 01 &  -1.077 &  \\
& 10 &  $13.68$ &  \\[4pt]

\multirow{3}{*}{1}
& 00 &  5.510   & \multirow{3}{*}{-0.553} \\
& 01 &  -5.266  &  \\
& 10 &  -2.164  &  \\[4pt]

\multirow{3}{*}{2}
& 00 &  -6.159  & \multirow{3}{*}{20.46} \\
& 01 &  1.775   &  \\
& 10 &  $23.78$  &  \\[4pt]

\multirow{3}{*}{3}
& 00 &  -1.751  & \multirow{3}{*}{-4.462} \\
& 01 &  3.437   &  \\
& 10 &  $-13.07$ &  \\[4pt]

\multirow{3}{*}{4}
& 00 &  -1.614  & \multirow{3}{*}{3.072} \\
& 01 &  -1.511  &  \\
& 10 &  $13.29$  &  \\[4pt]

\multirow{3}{*}{5}
& 00 &  2.595   & \multirow{3}{*}{3.251} \\
& 01 &  -0.399   &  \\
& 10 &  $-13.79$ &  \\
\bottomrule
\hline
\end{tabular}
\end{table}

\begin{table} 
\centering
\caption{Model fitting coefficients for $t^\omega_{\rm match}(q,e,l)$.}
\label{tab:tfreq-table}
\setlength{\tabcolsep}{26pt}   
\begin{tabular}{|c|c| r|}
\toprule
\hline
$k$ & $\alpha\beta$ & $A_{k\alpha\beta}\ (\times10^4)$\\
\hline
\midrule
\multirow{6}{*}{0}  
& 00 &  $25.41$ \\
& 01 & $-29.24$ \\
& 02 &  $1.710$  \\
& 10 & $-46.22$ \\
& 11 &  $70.79$ \\
& 20 & $-79.06$ \\[4pt]

\multirow{6}{*}{1}
& 00 &  $29.26$ \\
& 01 & $-36.35$  \\
& 02 &  $8.004$   \\
& 10 &  $26.74$  \\
& 11 & $-42.69$  \\
& 20 &  $64.40$  \\[4pt]

\multirow{6}{*}{2}
& 00 &  $30.87$ \\
& 01 & $-33.36$  \\
& 02 &  $1.900$  \\
& 10 & $-87.80$  \\
& 11 &  $95.12$   \\
& 20 & $-17.31$   \\[4pt]

\multirow{6}{*}{3}
& 00 &  $15.87$  \\
& 01 & $-18.34$   \\
& 02 &  $3.324$     \\
& 10 & $-23.32$   \\
& 11 &  $24.22$   \\
& 20 & $-23.03$   \\
\hline
\bottomrule
\end{tabular}
\end{table}


\bibliographystyle{apsrev4-1}
\bibliography{Manna_et_al_2024}
\end{document}